\documentclass[aps,12pt,showpacs,preprint,groupedaddress,floatfix]{revtex4}
\usepackage{graphicx}
\usepackage{dcolumn}
\usepackage{bm}
\usepackage{amssymb}
\usepackage{amsmath}

\begin{document}

\title{Microscopic Analysis of Nuclear Quantum Phase Transitions in 
the $\bm{ N \approx 90}$ region} 
\author{Z. P. Li}
\altaffiliation{School of Physics, Peking University, Beijing, China}
\author{T. Nik\v si\' c}
\author{D. Vretenar}
\affiliation{Physics Department, Faculty of Science, University of Zagreb,
Croatia}
\author{J. Meng}
\affiliation{School of Physics, Peking University, Beijing, China}
\author{G. A. Lalazissis}
\affiliation{Department of Theoretical Physics, Aristotle University of
Thessaloniki, GR-54124, Greece}
\author{P. Ring}
\affiliation{Physik-Department der Technischen Universit\"at M\"unchen,
D-85748 Garching,
Germany}
\date{\today}

\begin{abstract}
The analysis of shape transitions in Nd isotopes, based on the
framework of relativistic energy density functionals and restricted to axially 
symmetric shapes in Ref.~\cite{PRL99}, is extended to the 
region $Z = 60$, $62$, $64$ with $N \approx 90$, and includes 
both $\beta$ and $\gamma$ deformations. Collective excitation spectra
and transition probabilities are calculated starting from a five-dimensional 
Hamiltonian for quadrupole vibrational and rotational degrees of 
freedom, with parameters determined by constrained self-consistent relativistic 
mean-field calculations for triaxial shapes. The results reproduce 
available data, and show that there is an abrupt change of structure 
at $N=90$ that can be approximately characterized by the 
X(5) analytic solution at the critical point of the 
first-order quantum phase transition between spherical and 
axially deformed shapes. 
\end{abstract}

\maketitle

\section{\label{secI}Introduction}

The evolution of shell structures governs the variation of ground-state 
nuclear shapes along isotopic and isotonic chains. Nuclear structure explores 
a variety of phenomena related to structural evolution including, for instance,
the reduction of spherical shell gaps and modifications of magic numbers
in nuclei far from stability, occurrence of islands of inversion
and coexistence of shapes with different deformations. As the number of 
nucleons changes from nucleus to nucleus, in general one observes a 
gradual evolution of different shapes -- spherical, axially deformed, 
shapes that are soft with respect to triaxial deformations. These shape 
transitions reflect the underlying modifications of single-nucleon 
shell structure and interactions between valence nucleons. 
An especially interesting feature is the possible 
occurrence of shape phase transitions and critical-point phenomena 
for particular values of the number of protons and neutrons. 
Phase transitions in the equilibrium shapes of nuclei correspond 
to first- and second-order quantum phase transitions (QPT) between 
competing ground-state phases induced by variation of a non-thermal 
control parameter (number of nucleons) at zero 
temperature  \cite{Die.80,Feng.81,IZC.98}.

Nuclear quantum phase transitions have been the subject of extensive 
experimental and theoretical studies during the last decade. 
For recent reviews and an exhaustive bibliography we refer 
the reader to Refs.~\cite{JC.05,Rick.06,Rick.07,CJ.08,Bon.08}. Even 
though phase transitions in finite systems can only be defined in the 
classical limit in which the number of constituents tends to infinity, i.e. 
the transition is actually smoothed out in finite systems, there are nevertheless 
clear experimental signatures of abrupt changes in structure properties 
of finite nuclei with the addition or subtraction of only few nucleons. 
This is another distinct feature of QPT in atomic nuclei, i.e. the physical 
control parameter -- number of nucleons, can only take discrete integer 
values. Expectation values of suitably chosen operators, that as 
observables characterize the state of a nuclear system, can be used 
as order parameters \cite{IZ.04}. 

Theoretical studies of nuclear QPT are typically based on macroscopic 
geometric models of nuclear shapes and potentials \cite{Bon.08} and/or 
semi-microscopic algebraic models \cite{CJ.08}. 
In the geometric framework QPTs are analyzed in terms of a Bohr 
collective Hamiltonian for shape variables, and can be related to the concept of 
critical-point symmetries that provide parameter-independent 
predictions for excitation spectra and electromagnetic
transition rates for nuclei at the phase transition point. 
Alternatively, in the algebraic approach (e.g. the interacting boson model (IBM))  
different shapes coincide with particular dynamical symmetries of some algebraic 
structure, and the phase transition occurs when these symmetries are broken in 
a specific way. The two frameworks can be related, for instance, by using 
the coherent state formalism \cite{GK.80,DSI.80} which allows to establish a 
correspondence between the symmetry limits of the algebraic Hamiltonian and 
energy functionals expressed in terms of collective shape variables.
In both approaches, geometric and algebraic,  the description of QPT is 
based on model specific Hamiltonians which by 
construction describe shape changes. A shape phase transition is then 
accessed by variation of a control parameter. 

The two best studied classes of nuclear shape phase transitions, 
both theoretically and experimentally, correspond 
to a second-order QPT between spherical and $\gamma$-soft shapes 
\cite{FI.00,CZ.00}, and a first-order QPT between spherical and 
axially deformed shapes \cite{FI.01,CZ.01}. The former is a phase 
transition in one degree of freedom -- the axial deformation 
$\beta$ and, in the IBM language, represents a transition between the $U(5)$ and 
$O(6)$ dynamical symmetries in the limit of large boson number. 
The critical point of phase transition can also be related to a dynamical 
symmetry: in this case E(5) \cite{FI.00}, and the experimental realization of this 
critical-point symmetry was first identified in the spectrum of $^{134}$Ba \cite{CZ.00}. 
The second type of shape transitions, between spherical and axially deformed 
nuclei, is more commonly encountered and involves two degrees of 
freedom -- the deformations $\beta$ and $\gamma$.  The 
critical point of this phase transition, denoted X(5), does not correspond 
to a dynamical symmetry in the usual sense \cite{FI.01}. Nevertheless, 
for the particular case in which the $\beta$ and $\gamma$ degrees of freedom 
are decoupled, an approximate analytic solution at the critical point of phase 
transition  can be expressed in terms of zeros of Bessel functions of irrational 
order \cite{FI.01}. Evidence for the empirical realization of X(5) 
phase transition was first reported for $^{152}$Sm and other $N=90$ isotones 
in Ref.~\cite{CZ.01}.

Even though phenomenological approaches to nuclear QPT have been very 
successful, and the predicted isotopic trends for various observables are in 
very good agreement with data, it would clearly be desirable to have a 
fully microscopic description of shape phase transitions, starting from 
nucleonic degrees of freedom. This is especially important in view of 
the fact that the physical control parameter in nuclear QPT is the actual
number of nucleons in a nucleus, rather than a strength parameter of a 
model specific, Ising-type Hamiltonian. 

Several recent studies \cite{Meng.05,SG.05,FBL.06,RS.07,RE.08,RRS.08} 
have reported microscopic calculations of potential 
energy surfaces as functions of quadrupole deformation, for a 
number of isotopic chains in which the occurrence of shape phase
transitions had been predicted. Potentials calculated 
with a constraint on the mass quadrupole moment display shape 
transitions from spherical to deformed configurations. 
In Ref.~\cite{PRL99} we analyzed shape transitions in Nd isotopes, 
using a microscopic model based on relativistic energy density functionals (EDF).
Starting from constrained self-consistent mean-field 
calculations of potential energy curves,  the generator
coordinate method (GCM)~\cite{RS.80} was used to perform configuration
mixing of angular-momentum and particle-number projected relativistic
wave functions. It was shown that the microscopic framework based on
universal EDFs, adjusted to nuclear ground-state properties, and
extended to take into account correlations related to symmetry
restoration and fluctuations of collective variables, describes not
only general features of shape transitions, but also the unique
behavior of excitation spectra and transition rates at the
critical point of quantum shape phase transition. In particular, 
the particle-number projected GCM spectra, intraband and 
interband B(E2) values for $^{148,150,152}$Nd were found in 
excellent agreement with data, and close to the X(5)-model 
predictions for $^{150}$Nd. The self-consistent GCM calculation based 
on the relativistic density functional PC-F1~\cite{BMM.02}, predicted the 
shape phase transition precisely at the isotope $^{150}$Nd, 
in agreement with empirical evidence for the realization of 
X(5) in the $N=90$ rare-earth isotones \cite{Rick.06,Rick.07}.

X(5) denotes a particular model for a first-order QPT between 
spherical and axially deformed shapes, based on the assumption of a 
separable potential $V(\beta,\gamma) = V_{\beta} (\beta) + V_{\gamma}(\gamma)$. 
By neglecting the small barrier between the competing spherical and 
deformed minima, i.e. taking an infinite square well in the variable $\beta$,  
and assuming an axially stabilized potential $ V_{\gamma}(\gamma)$, 
an approximate analytic solution is obtained at the point of phase 
transition \cite{FI.01}. The microscopic axially symmetric potential for $^{150}$Nd, 
calculated with self-consistent mean-field models
(cf. Fig. 1 in Ref.~\cite{PRL99}) is, of course, more realistic 
than the infinite square well considered by Iachello in Ref.~\cite{FI.01}. 
First, although it displays a wide flat minimum in $\beta$ on the prolate side, 
the flat bottom of the potential does not start at $\beta = 0$, i.e. the 
coexisting shapes do not include the spherical configuration. In this 
respect the microscopic potential is closer to the generalization 
of the axially symmetric X(5) solution for the collective Bohr Hamiltonian 
to the transition path between X(5) and the rigid-rotor 
limit, represented by infinite square-well potentials over a confined range 
of values $\beta_M > \beta_m \geq 0$ (confined $\beta$-soft potential, 
$\beta_m$ and $\beta_M$ denote the positions of the inner and outer 
walls on the $\beta$ axis, respectively) \cite{PG.04}.
Second, the microscopic potential well has softer walls compared to the  
X(5) square well. The unrealistic features introduced by the hard wall of 
the square well were analyzed in Ref.~\cite{caprio}, and it was noted that 
the compression of the wave function against the well wall induces effects 
in spectra approximating rigid  $\beta$ deformation. 

More importantly, the GCM calculation of Ref.~\cite{PRL99} was restricted to 
axially symmetric shapes ($\gamma = 0$). We note that an exactly separable 
$\gamma$-rigid version (with $\gamma = 0$) of the X(5) model has 
been constructed in Ref.~\cite{Bon.06}. Although the original X(5) solution 
relied on an approximate separation of the potential in the variables
$\beta$ and $\gamma$ \cite{FI.01}, the exact diagonalization of the 
Bohr  Hamiltonian with a $\beta$-soft, axially stabilized potential, carried 
out in Ref.~\cite{caprio}, has shown that many properties of the solution are 
dominated by $\beta - \gamma$ coupling induced by the kinetic energy 
operator. Band-head excitation energies, energy spacings within the bands, 
and transition strengths are strongly dependent on the $\gamma$ stiffness of the 
potential. The importance of the explicit treatment of the triaxial degree of 
freedom, i.e. inclusion of $\beta - \gamma$ coupling, was also emphasized 
in two recent studies \cite{RE.08,RRS.08} that used the self-consistent 
Hartree-Fock-Bogoliubov model, based on the finite-range and density-dependent 
Gogny interaction, to generate potential energy surfaces in the $\beta - \gamma$ 
plane. In addition, in Ref.~\cite{RE.08} excitation spectra of Nd isotopes were 
calculated using the GCM with particle-number and angular-momentum 
projected axially symmetric wave functions. However, while 
GCM configuration mixing of axially symmetric mean-field states is
routinely performed in studies of collective excitation spectra, 
the application of this method to triaxial shapes
is a much more difficult problem. Only very recently a model has been
introduced~\cite{BH.08}, based on triaxial mean-field states, 
projected on particle number and angular momentum 
and mixed by the generator coordinate method. 
The numerical implementation of the model is rather complex, and applications to
medium-heavy and heavy nuclei are still computationally too demanding. 
An alternative approach is the explicit solution of the collective 
Hamiltonian in five dimensions, with deformation-dependent parameters 
determined from microscopic self-consistent mean-field calculations. This approach 
will be used in the present study. 

In this work we therefore extend the analysis of shape transitions in Nd isotopes 
of Ref.~\cite{PRL99}, to the region $Z = 60$, $62$, $64$ with $N \approx 90$, 
and include the explicit treatment of both $\beta$ and $\gamma$ degrees of 
freedom. Collective excitation spectra and transition probabilities are calculated 
starting from a five-dimensional collective Hamiltonian for quadrupole vibrational 
and rotational degrees of freedom, with parameters determined by 
constrained self-consistent relativistic mean-field calculations for triaxial shapes.

An outline of the theoretical framework is included in Sec.~\ref{secII}: the 
method of solution of the eigenvalue problem of
the general collective Hamiltonian, and the calculation of the mass
parameters, moments of inertia, and zero-point energy corrections. 
In Sec.~\ref{secIII} we present a detailed study of shape transitions 
in the  $N \approx 90$ rare-earth nuclei. Theoretical results are compared with 
experimental excitation spectra and transition rates, as well as 
with the X(5) approximate solution at  the point of first-order quantum phase 
transition. Sec.~\ref{secIV} presents a summary and an outlook for future 
studies. 
\section{\label{secII} Collective Hamiltonian in Five Dimensions}

In Ref.~\cite{NikLi} we have developed a model for
the solution of the eigenvalue problem of a five-dimensional
collective Hamiltonian for quadrupole vibrational and rotational
degrees of freedom, with parameters determined by constrained
self-consistent relativistic mean-field calculations for triaxial
shapes. The five quadrupole collective coordinates 
are parameterized in terms of two
deformation parameters $\beta$ and $\gamma$, and three Euler angles
$(\phi,\;\theta,\;\psi)\equiv \Omega$, which define the orientation
of the intrinsic principal axes in the laboratory frame. 
The collective Hamiltonian can be written in the form:
\begin{equation}
\label{hamiltonian-quant}
\hat{H} = \hat{T}_{\textnormal{vib}}+\hat{T}_{\textnormal{rot}}
              +V_{\textnormal{coll}} \; ,
\end{equation}
with the vibrational kinetic energy:
\begin{align}
\hat{T}_{\textnormal{vib}} =&-\frac{\hbar^2}{2\sqrt{wr}}
   \left\{\frac{1}{\beta^4}
   \left[\frac{\partial}{\partial\beta}\sqrt{\frac{r}{w}}\beta^4
   B_{\gamma\gamma} \frac{\partial}{\partial\beta}
   - \frac{\partial}{\partial\beta}\sqrt{\frac{r}{w}}\beta^3
   B_{\beta\gamma}\frac{\partial}{\partial\gamma}
   \right]\right.
   \nonumber \\
   &+\frac{1}{\beta\sin{3\gamma}}\left.\left[
   -\frac{\partial}{\partial\gamma} \sqrt{\frac{r}{w}}\sin{3\gamma}
      B_{\beta \gamma}\frac{\partial}{\partial\beta}
    +\frac{1}{\beta}\frac{\partial}{\partial\gamma} \sqrt{\frac{r}{w}}\sin{3\gamma}
      B_{\beta \beta}\frac{\partial}{\partial\gamma}
   \right]\right\} \; ,
\end{align}
and rotational kinetic energy:
\begin{equation}
\hat{T}_{\textnormal{\textnormal{\textnormal{rot}}}} = 
\frac{1}{2}\sum_{k=1}^3{\frac{\hat{J}^2_k}{\mathcal{I}_k}} \; .
\end{equation}
$V_{\textnormal{coll}}$ is the collective potential. 
$\hat{J}_k$ denotes the components of the angular momentum in
the body-fixed frame of a nucleus, and the mass parameters
$B_{\beta\beta}$, $B_{\beta\gamma}$, $B_{\gamma\gamma}$, as well as 
the moments of inertia $\mathcal{I}_k$, depend on the quadrupole
deformation variables $\beta$ and $\gamma$:
\begin{equation}
\mathcal{I}_k = 4B_k\beta^2\sin^2(\gamma-2k\pi/3) \;.
\end{equation}
Two additional quantities that appear in the expression for the vibrational energy:
$r=B_1B_2B_3$, and $w=B_{\beta\beta}B_{\gamma\gamma}-B_{\beta\gamma}^2 $, 
determine the volume element in the collective space.
The Hamiltonian Eq.~(\ref{hamiltonian-quant}) describes quadrupole vibrations,
rotations, and the coupling of these collective modes. The corresponding 
eigenvalue problem is solved using an expansion of eigenfunctions in terms 
of a complete set of basis
functions that depend on the deformation variables $\beta$ and
$\gamma$, and the Euler angles $\phi$, $\theta$ and $\psi$ \cite{Pro.99}. 

The dynamics of the collective Hamiltonian is governed by the
seven functions of the intrinsic deformations $\beta$ and $\gamma$:
the collective potential, the three mass parameters:
$B_{\beta\beta}$, $B_{\beta\gamma}$, $B_{\gamma\gamma}$, and the
three moments of inertia $\mathcal{I}_k$. These functions are
determined by the choice of a particular microscopic nuclear energy
density functional or effective interaction. As in our previous 
study of axial shape transitions in Nd isotopes~\cite{PRL99}, 
also in this work we use the relativistic functional PC-F1 (point-coupling
Lagrangian)~\cite{BMM.02} in the particle-hole channel, and a
density-independent $\delta$-force is the effective interaction in
the particle-particle channel. The parameters of the PC-F1 functional
and the pairing strength constants $V_n$ and $V_p$ have been adjusted
simultaneously to
ground-state observables (binding energies, charge and diffraction
radii, surface thickness and pairing gaps) of spherical nuclei
~\cite{BMM.02}, with pairing correlations treated in the BCS
approximation.

The map of the energy surface as function of the quadrupole
deformation is obtained by imposing constraints on the axial and
triaxial mass quadrupole moments. The method of quadratic constraints
uses an unrestricted variation of the function
\begin{equation}
\langle H\rangle
   +\sum_{\mu=0,2}{C_{2\mu}\left(\langle \hat{Q}_{2\mu}  \rangle - q_{2\mu}  \right)^2} \; ,
\label{constr}
\end{equation}
where $\langle H\rangle$ is the total energy, and 
$\langle \hat{Q}_{2\mu}\rangle$ denotes the expectation value of the mass quadrupole 
operator:
\begin{equation}
\hat{Q}_{20}=2z^2-x^2-y^2 \quad \textnormal{and}\quad \hat{Q}_{22}=x^2-y^2 \;.
\end{equation}
$q_{2\mu}$ is the constrained value of the multipole moment,
and $C_{2\mu}$ the corresponding stiffness constant~\cite{RS.80}.

The moments of inertia are calculated according to the Inglis-Belyaev
formula:~\cite{Ing.56,Bel.61}
\begin{equation}
\label{Inglis-Belyaev}
\mathcal{I}_k = \sum_{i,j}{\frac{\left(u_iv_j-v_iu_j \right)^2}{E_i+E_j}
  \langle i |\hat{J}_k | j  \rangle |^2}\quad k=1,2,3,
\end{equation}
where $k$ denotes the axis of rotation, and the summation runs over
the proton and neutron quasiparticle states. The quasiparticle
energies $E_i$, occupation probabilities $v_i$, and single-nucleon
wave functions $\psi_i$ are determined by solutions of the
constrained relativistic mean-field (RMF) plus BCS equations. 
The mass parameters associated with
the two quadrupole collective coordinates
$q_0=\langle\hat{Q}_{20}\rangle$ and $q_2=\langle\hat{Q}_{22}\rangle$
are also calculated in the cranking approximation~\cite{GG.79}
\begin{equation}
\label{masspar-B}
B_{\mu\nu}(q_0,q_2)=\frac{\hbar^2}{2}
 \left[\mathcal{M}_{(1)}^{-1} \mathcal{M}_{(3)} \mathcal{M}_{(1)}^{-1}\right]_{\mu\nu}\;,
\end{equation}
with
\begin{equation}
\label{masspar-M}
\mathcal{M}_{(n),\mu\nu}(q_0,q_2)=\sum_{i,j}
 {\frac{\left\langle i\right|\hat{Q}_{2\mu}\left| j\right\rangle
 \left\langle j\right|\hat{Q}_{2\nu}\left| i\right\rangle}
 {(E_i+E_j)^n}\left(u_i v_j+ v_i u_j \right)^2}\;.
\end{equation}

The collective energy surface includes the energy of zero-point
motion (ZPE), that corresponds to a superposition of zero-point motion 
of individual nucleons in the single-nucleon potential.  The ZPE correction  
depends on the deformation, and includes terms originating from the vibrational
and rotational kinetic energy, and a contribution of potential energy
\begin{equation}
\Delta V(q_0,q_2)=\Delta V_{\textnormal{vib}}(q_0,q_2) 
                            + \Delta V_{\textnormal{rot}}(q_0,q_2)
                            + \Delta V_{\textnormal{pot}}(q_0,q_2) \; .
\end{equation}
The latter is much smaller than the kinetic energy contribution,
and is usually neglected. The vibrational and rotational ZPE 
are calculated in the cranking approximation \cite{GG.79}, i.e. on the 
same level of approximation as the mass
parameters and the moments of inertia. The potential 
$V_{\textnormal{coll}}$ in the collective Hamiltonian
Eq.~(\ref{hamiltonian-quant}) is obtained by subtracting the ZPE
corrections from the total energy that corresponds to the
solution of constrained RMF+BCS equations (cf. Eq~(\ref{constr}) ), 
at each point on the triaxial deformation plane:
\begin{equation}
\label{Vcoll}
{V}_{\textnormal{coll}}(q_0,q_2) = E_{\textnormal{tot}}(q_0,q_2)
  - \Delta V_{\textnormal{vib}}(q_0,q_2) - \Delta V_{\textnormal{rot}}(q_0,q_2) \; .
\end{equation}
\section{\label{secIII}Shape Phase Transitions in the 
$\bm{ N\approx 90}$ region}

The present study considers the Nd, Sm and Gd nuclei with 
$N\approx 90$. These isotopic chains display transitions 
from spherical to deformed shapes, and experimental evidence 
has recently been reported \cite{Rei.02,CZ.01,Ton.04} for the $N=90$ 
isotones: $^{150}$Nd, $^{152}$Sm, $^{154}$Gd lying close to the 
point of first-order QPT.
We will first extend our analysis of Ref.~\cite{PRL99} of the Nd isotopic 
chain to explicitly include the triaxial degree of freedom. The microscopic 
picture of shape phase transitions at $N=90$ will be analyzed in detail for the 
Nd nuclei, and in the remainder of this section the most interesting results 
for the Sm and Gd isotopic chains will be presented.

\begin{figure}
\includegraphics[scale=1.0]{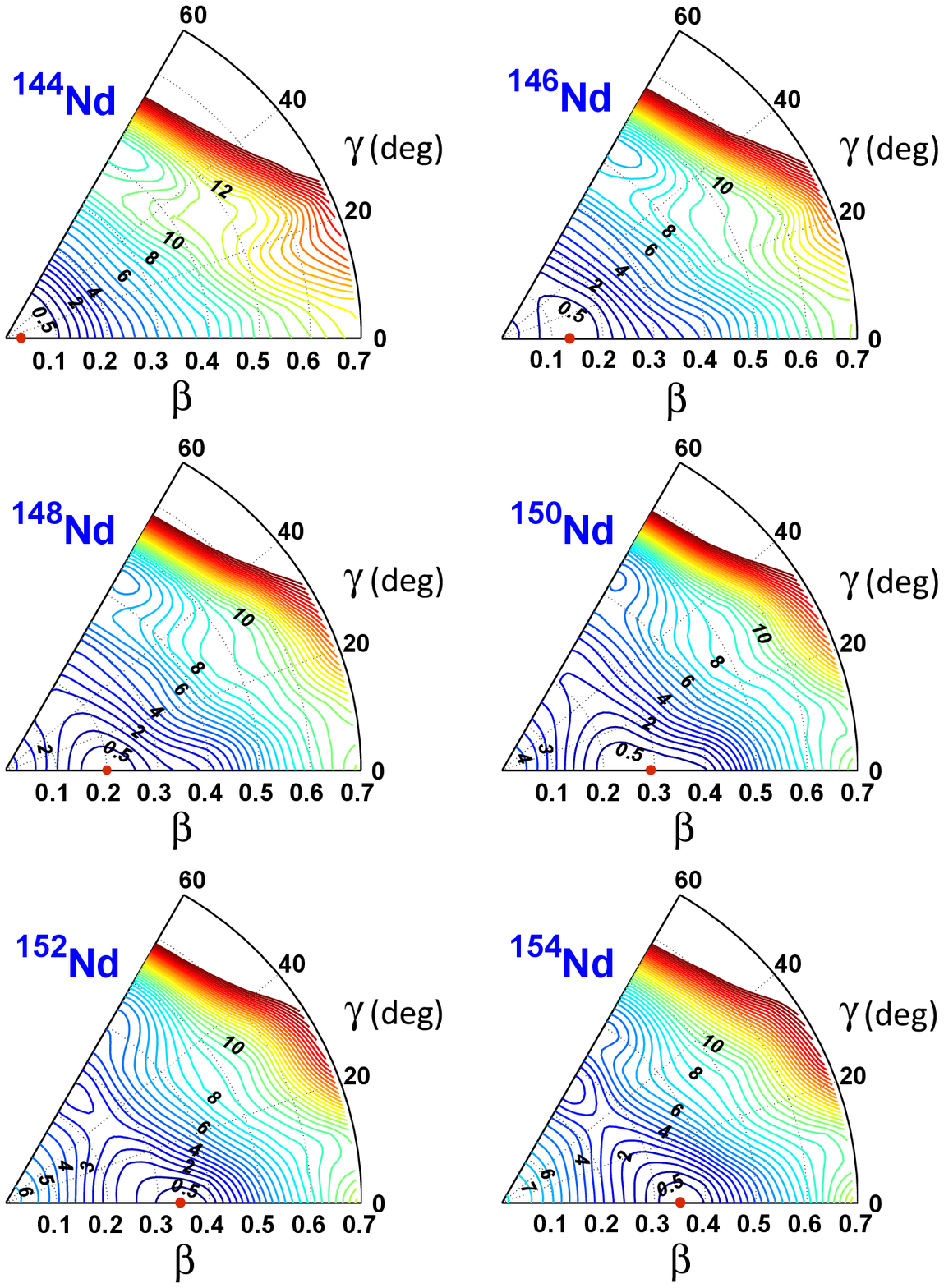}
\caption{(Color online) Self-consistent RMF+BCS triaxial quadrupole
binding energy maps of the even-even $^{144-154}$Nd isotopes
in the $\beta - \gamma$ plane ($0\le \gamma\le 60^0$).
All energies are normalized with respect to
the binding energy of the absolute minimum (red filled circle). 
The contours join points
on the surface with the same energy (in MeV).}
\label{FigA}
\end{figure}

\begin{figure}
\includegraphics[scale=1.2,angle=0]{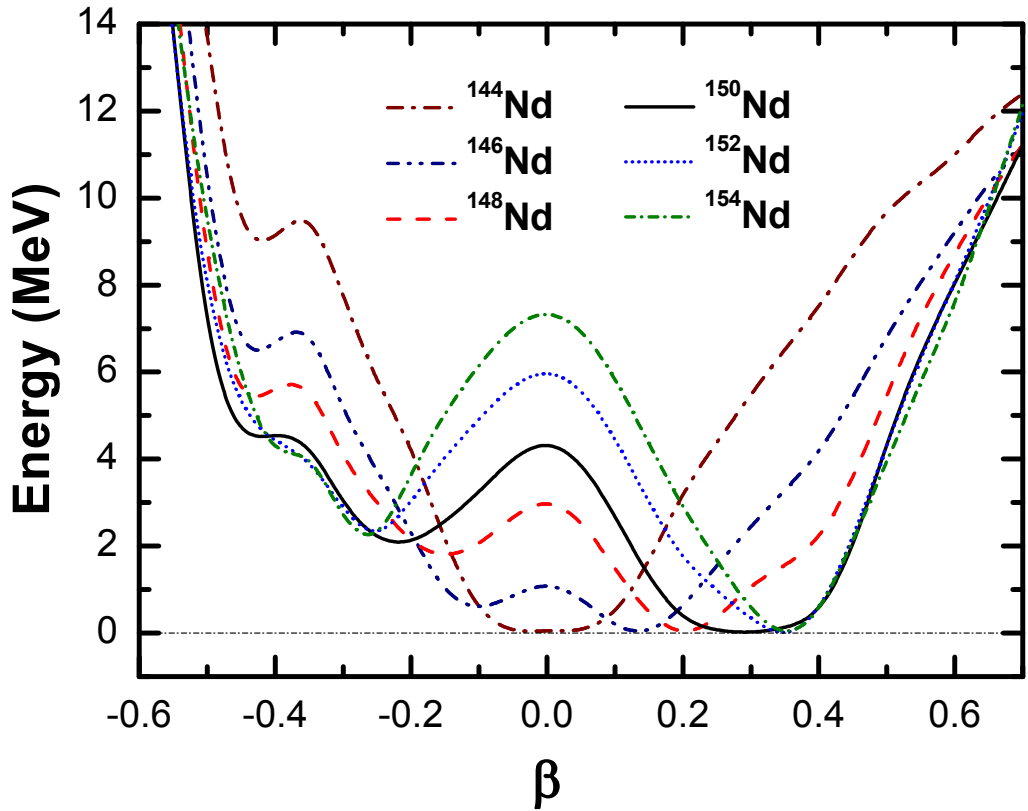}
\caption{(Color online)
Self-consistent RMF+BCS binding energy curves of the even-even $^{144-154}$Nd
isotopes, as functions of the axial deformation parameter $\beta$. 
Energies are normalized with respect to
the binding energy of the absolute minimum for a given isotope.
Negative values of $\beta$ correspond to the oblate 
$(\beta>0,\;\gamma=60^0)$ axis on the
$\beta - \gamma$ plane.}
\label{FigB}
\end{figure}

\begin{figure}
\includegraphics[scale=1.2,angle=0]{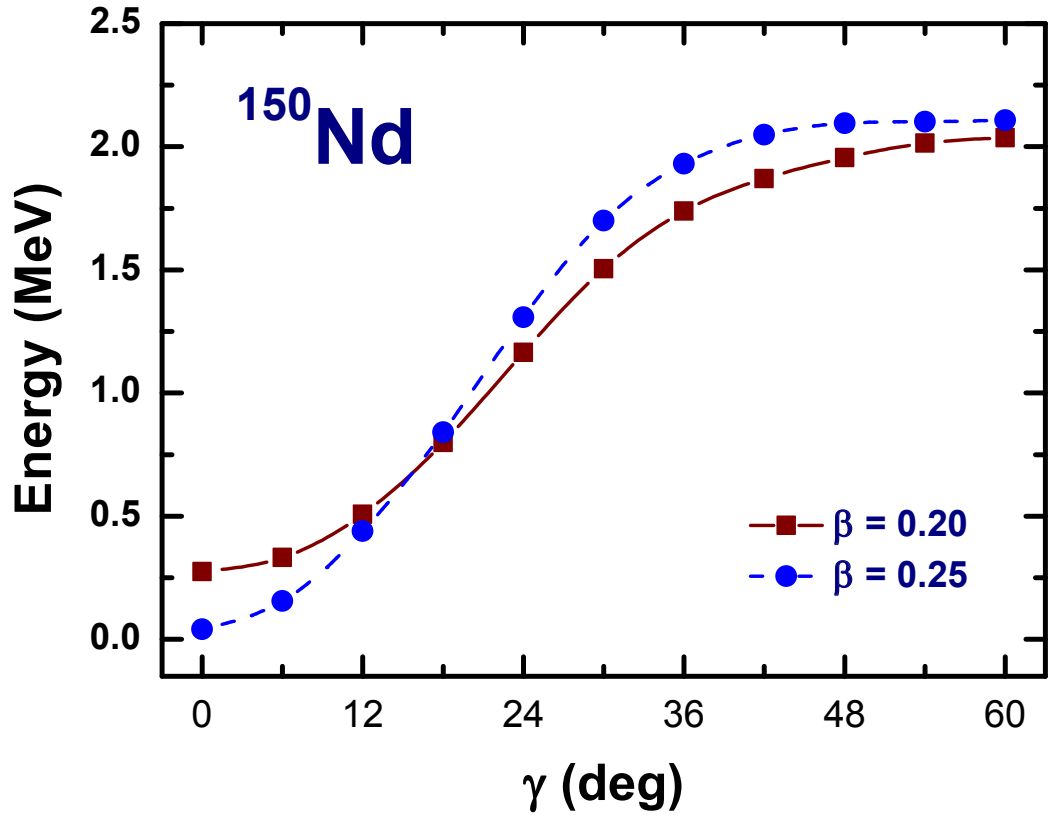}
\caption{(Color online)
Self-consistent RMF+BCS binding energy curves of the $^{150}$Nd
nucleus, as functions of the deformation parameter $\gamma$, 
for two values of the axial deformation $\beta =0.2$ and $0.25$.}
\label{FigC}
\end{figure}

In  Fig.~\ref{FigA} we display the self-consistent
RMF+BCS triaxial quadrupole binding energy maps of
the even-even $^{144-154}$Nd in
the $\beta - \gamma$ plane
 ($0\le \gamma\le 60^0$), obtained by imposing constraints on expectation
values of the quadrupole moments $\langle \hat{Q}_{2 0}  \rangle$ and
$\langle \hat{Q}_{2 2}  \rangle$ (cf. Eq.~(\ref{constr})). Filled circle symbols
denote absolute minima; all energies are normalized with respect to
the binding energy of the absolute minimum. The contours join points
on the surface with the same energy. The energy maps nicely illustrate the
gradual increase of deformation of the prolate minimum with increasing
number of neutrons, from the spherical $^{144}$Nd to the strongly deformed 
$^{154}$Nd, and the evolution of the $\gamma$-dependence of 
the potentials. With increasing neutron number the Nd isotopes remain 
prolate deformed in the lowest state, i.e. the shape evolution corresponds, 
in the language of the interacting boson model, to a 
transition between the U(5) and SU(3) limits of the Casten symmetry 
triangle \cite{Rick.06,Rick.07}. One notes, however, the appearance 
of oblate minima in $^{152}$Nd and $^{154}$Nd. An important feature 
of the energy maps shown in Fig.~\ref{FigA} is the extended flat minimum 
around $\beta \approx 0.3$ in $^{150}$Nd. This is seen even more clearly 
in the axial maps of Fig.~\ref{FigB}, where we plot the binding energies 
as functions of the axial deformation parameter $\beta$. 
Energies are normalized with respect to
the binding energy of the absolute minimum for a given isotope.
Negative values of $\beta$ correspond to the oblate 
$(\beta>0,\;\gamma=60^0)$ axis on the
$\beta - \gamma$ plane. The prolate minimum gradually shifts to 
larger deformation and saturates at $\beta \approx 0.35$ in 
$^{152}$Nd and $^{154}$Nd. It is interesting that for these two nuclei 
the calculation also predicts the same deformation $\beta \approx -0.3$ 
of the oblate minima. The difference is in the barriers separating the 
prolate and oblate minima: 6 MeV for $^{152}$Nd, and more than 7 MeV 
for $^{154}$Nd. For $^{150}$Nd, on the other hand, the flat prolate 
minimum extends in the interval $0.2 \leq \beta \leq 0.4$, whereas 
the oblate minimum at  $\beta \approx -0.2$ is not a 
true minimum, but rather a saddle point in the $\beta - \gamma$ plane 
(cf. Fig.~\ref{FigA}). This is illustrated in Fig.~\ref{FigC}, 
where we plot  the binding energy curves of $^{150}$Nd
as functions of the deformation parameter $\gamma$, 
for two values of the axial deformation $\beta =0.2$ and $0.25$. 
In the region of the flat prolate minimum the potential displays a 
parabolic dependence on $\gamma$ for $\gamma \leq 30^\circ$,
but flattens with increasing $\gamma$ because, of course, it must be 
periodic with a period $2 \pi/3$. The difference between 
the two curves plotted in Fig.~\ref{FigC} shows that the $\gamma$-potential 
is not completely independent of $\beta$, as it was originally 
assumed for the X(5) model in Ref.~\cite{FI.01}. We note that both in the 
test of the X(5) for the $\gamma$ degree of freedom of Ref.~\cite{BCZ.03}, 
as well as in the study of the validity of the approximate separation of 
variables introduced with the X(5) model \cite{caprio}, the potential 
in $\gamma$ was chosen to be a harmonic oscillator potential.

Potential energy surfaces similar to those shown in Fig.~\ref{FigA} 
were also obtained in two recent studies \cite{RE.08,RRS.08} that used the 
self-consistent Hartree-Fock-Bogoliubov (HFB) model, based on the finite-range 
and density-dependent Gogny interaction. However, compared to the 
results shown in Figs.~\ref{FigA} and \ref{FigB}, the flat 
prolate minimum in $^{150}$Nd is less pronounced in the HFB calculation 
with the Gogny force. By performing the corresponding GCM calculation 
with particle-number and angular-momentum projected axially symmetric 
wave functions, the authors of Ref.~\cite{RE.08} show that, rather 
than $^{150}$Nd,  the excitation energies and transition
probabilities for $^{148}$Nd are actually closer to the X(5) model predictions.  
In fact, they notice that intrinsic potentials rather different 
from the one considered in the original X(5) model \cite{FI.01} 
(wide flat minimum in $\beta$ starting at $\beta = 0$), lead to
excitation spectra and transition rates similar to the X(5) predictions, 
and thus question the use of the axial deformation $\beta$ as order 
parameter of a shape phase transition.

Starting from constrained self-consistent solutions, i.e. using 
single-particle wave functions, occupation probabilities, and
quasiparticle energies that correspond to each point on the energy
surfaces shown in Figs.~\ref{FigA}, the parameters that
determine the collective Hamiltonian: mass parameters
$B_{\beta\beta}$, $B_{\beta\gamma}$, $B_{\gamma\gamma}$, three
moments of inertia $\mathcal{I}_k$, as well as the zero-point energy
corrections, are calculated as functions of the deformations $\beta$
and  $\gamma$. The diagonalization of the resulting Hamiltonian yields the
excitation energies $E_\alpha^I$ and the collective wave functions:
\begin{equation}
\label{wave-coll}
\Psi_\alpha^{IM}(\beta,\gamma,\Omega) =
  \sum_{K\in \Delta I}
           {\psi_{\alpha K}^I(\beta,\gamma)\Phi_{MK}^I(\Omega)}.
\end{equation}
The angular part corresponds to linear combinations of the Wigner
functions
\begin{equation}
\label{Wigner}
\Phi_{MK}^I(\Omega)=\sqrt{\frac{2I+1}{16\pi^2(1+\delta_{K0})}}
\left[D_{MK}^{I*}(\Omega)+(-1)^ID_{M-K}^{I*}(\Omega) \right] \; ,
\end{equation}
and the summation in Eq. (\ref{wave-coll}) is over the allowed set  of
the $K$ values:
\begin{equation}
\Delta I = \left\{ \begin{array}{c}
   0,2,\dots,I \quad \textnormal{for} \quad  I\; \textnormal{mod}\; 2 = 0 \\
   2,4,\dots,I-1 \quad \textnormal{for} \quad   I\; \textnormal{mod}\; 2 =1
\end{array} \right .
\end{equation} 
In addition to the
yrast ground-state band, in deformed and transitional nuclei excited
states are usually also assigned to (quasi) $\beta$ and $\gamma$
bands. This is done according to the distribution of the angular
momentum projection $K$ quantum number.
Excited states with predominant $K=2$ components in the wave function
are assigned to the $\gamma$-band, whereas the $\beta$-band comprises
states above the yrast characterized by dominant $K=0$ components. 
The mixing of different intrinsic configurations in the state
$|\alpha I\rangle$ can be determined from
the distribution of the projection $K$ of the angular momentum $I$
on the $z$ axis in the body-fixed frame:
\begin{equation}
N_K=6\int_0^{\pi/3}{\int_0^\infty{
   |\psi^I_{\alpha,K}(\beta,\gamma)|^2\beta^4|\sin{3\gamma}|d\beta d\gamma}},
\label{NK}
\end{equation}
where the components $\psi^I_{\alpha,K}(\beta,\gamma)$ are defined in 
Eq.~(\ref{wave-coll}).
When $K$ is a good quantum number, only one of the integrals Eq.~(\ref{NK})
will give a value close to $1$. A broader distribution of $N_K$
values in the state $|\alpha I\rangle$ provides a measure of mixing
of intrinsic configurations. 

\begin{figure}
\includegraphics[scale=0.55,angle=0]{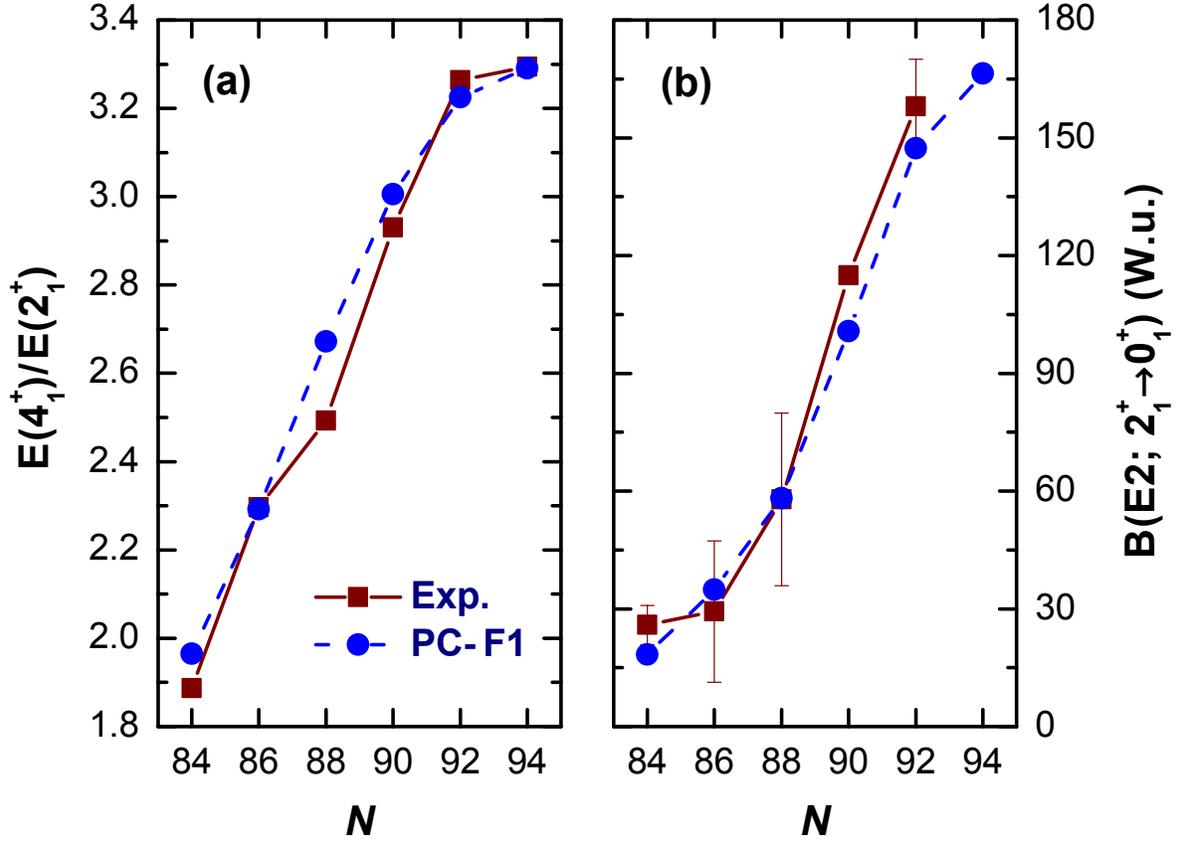}
\caption{(Color online)
Evolution of the characteristic collective observables $R_{4/2}$ and 
B(E2; $2^+_1 \to 0^+_1$) (in Weisskopf units) with neutron number in 
Nd isotopes. The microscopic values calculated with the PC-F1 energy 
density functional are shown in comparison with data \cite{Rei.02,NNDC,LBNL}.}
\label{FigD}
\end{figure}

Before comparing the calculated excitation spectrum of  $^{150}$Nd 
with data and X(5) model predictions, we will consider some characteristic 
signatures of shape transitions in Nd isotopes. One of the distinct 
features of shape transitions is the ratio between the excitation energies 
of the first $4^+$ and $2^+$ states along an isotopic chain. For a 
transition between spherical and axially deformed shapes, this
quantity varies from the value $R_{4/2} \equiv E(4_1^+)/E(2_1^+) = 2$ 
characteristic for a spherical vibrator (U(5) symmetry limit), 
to $R_{4/2} = 3.33$ for a well deformed axial rotor (SU(3) symmetry 
limit). In the left panel of Fig.~\ref{FigD} we plot the theoretical values of 
$R_{4/2}$ for the six Nd isotopes, calculated from the spectrum of 
the collective Hamiltonian with mass parameters, moments of inertia 
and potentials determined by the PC-F1 density functional, in comparison with 
experimental values. The calculation reproduces in detail the rapid increase 
of $R_{4/2}$ from the spherical value of $\approx 1.9$ in $^{144}$Nd, to 
$\approx 3.3$ characteristic for the well deformed rotors $^{152}$Nd and 
$^{154}$Nd. For the $N=90$ isotope, in particular, the calculated 
$R_{4/2} = 3$ and experimental $R_{4/2} = 2.93$ values, are very close to 
the characteristic, parameter-free, X(5) model prediction 
$R_{4/2} = 2.91$ \cite{FI.01}. Note, however, that our microscopic results for 
$R_{4/2}$, as well as other characteristic quantities, include the effect 
of $\beta - \gamma$ coupling, absent in the original X(5)-symmetry model. 
The panel on the right of Fig.~\ref{FigD} illustrates the evolution of another 
characteristic collective observable with neutron number:
B(E2; $2^+_1 \to 0^+_1$) (in Weisskopf units). 
The important result here is not only that the calculation reproduces 
the swift increase of the empirical B(E2) values from less than 30  
Weisskopf units in $^{144}$Nd, to more than 160 Weisskopf units in 
$^{152}$Nd, but also that the calculation is completely parameter-free.
Namely, an important advantage of using structure models based on
self-consistent mean-field single-particle solutions is the fact that
physical observables, such as transition probabilities and
spectroscopic quadrupole moments, are calculated in the full
configuration space and there is no need for effective charges. Using
the bare value of the proton charge in the electric quadrupole
operator $\mathcal{\hat{M}}(E2)$, the transition probabilities between
eigenstates of the collective Hamiltonian can be directly compared
with data. This is in contrast to, for instance, algebraic (interacting boson model)  
or shell-model approaches, that explicitly consider only valence nucleons
and, therefore, calculate transition rates by adjusting the effective 
charges to reproduce some empirical values. For example, 
the X(5) predictions for transition strengths are normalized
to the experimental B(E2; $2^+_1 \to 0^+_1$). We also note 
that the present results, both for $R_{4/2}$ and 
B(E2; $2^+_1 \to 0^+_1$), are in better agreement with data 
than those obtained in the axial GCM calculation with the 
Gogny force in  Ref.~\cite{RE.08}. In particular, the transition from 
spherical to well deformed Nd nuclei is less abrupt than in the 
axial GCM calculation and, therefore, the solutions of the collective 
Hamiltonian in five dimensions including $\beta - \gamma$ coupling, 
describe much better the transitional nuclei $^{148}$Nd and 
$^{150}$Nd.

\begin{figure}
\includegraphics[scale=0.55,angle=0]{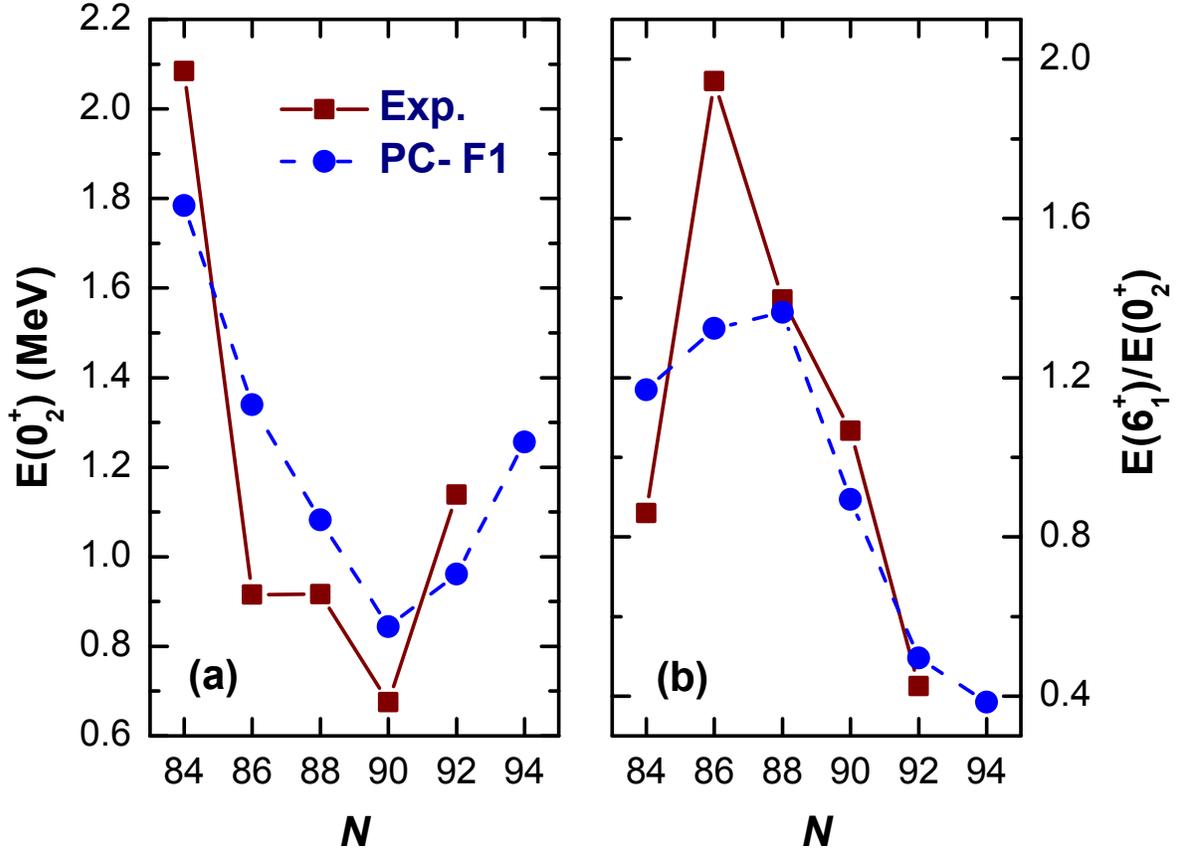}
\caption{(Color online)
Evolution of the first excited $0^+$ state and the ratio $E(6^+_1)/E(0^+_2)$ 
with neutron number in Nd isotopes. The microscopic values calculated with 
the PC-F1 energy density functional are compared to data \cite{NNDC,LBNL}.}
\label{FigE}
\end{figure}

In Fig.~\ref{FigE} we display the isotopic dependence 
of the first excited $0^+$ state, and the ratio $E(6^+_1)/E(0^+_2)$ 
in Nd nuclei. The microscopic excitation energies calculated with 
the PC-F1 energy density functional are compared with experimental 
values. The calculation reproduces the empirical trend and, in particular, 
it predicts that the first excited $0^+$ state has the lowest excitation 
energy at $N=90$, in agreement with data. With the exception of the 
very low $0^+_2$ state in $^{146}$Nd, the calculated excitation energies 
are also in quantitative agreement with experimental values. 
The fact that the band-head $0^+_2$ of the quasi-$\beta$ band has the 
lowest excitation energy in $^{150}$Nd can be attributed to the softness 
(flatness) of the potential with respect to $\beta$ deformation (cf. Figs. 
\ref{FigA} and \ref{FigB}). With the increase of the stiffness of the 
potential in $^{152}$Nd and $^{154}$Nd, the position of the $\beta$-band is 
shifted to higher excitation energies, and the ratio $E(6^+_1)/E(0^+_2)$ 
decreases to less than 1/2.

The prediction of a near degeneracy of the $6_1^+$ level of the ground state 
band and the first excited $0^+$ state is another key signature of the X(5) model 
\cite{Rick.06,Rick.07}. In a very recent study of empirical order parameters for 
first-order nuclear QPT \cite{Dennis.08}, Bonatsos {\it et al.} 
have shown that the ratio $E(6^+_1)/E(0^+_2)$ presents an effective 
order parameter of a first-order phase transition, and takes the special value 
of $\approx 1$ close to the phase transition point. Furthermore, results of 
extensive IBM calculations with large boson number $N_B$ show that the 
degeneracy of $E(0^+_2)$ and $E(6^+_1)$ is a signature of a {\em line} 
of first-order transitions that extends across the Casten symmetry triangle. 
The ratios $E(6^+_1)/E(0^+_2)$ obtained with the PC-F1 energy density 
functional are compared to the experimental values in Fig.~\ref{FigE}(b). 
We notice that the
calculation reproduces the abrupt decrease of this quantity between 
$^{148}$Nd and $^{152}$Nd, with a value close to 1 in $^{150}$Nd. 
The isotopic dependence of the theoretical values corresponds to the 
one predicted for a first-order phase transition in Ref.~\cite{Dennis.08}.

\begin{figure}
\includegraphics[scale=0.75,angle=270]{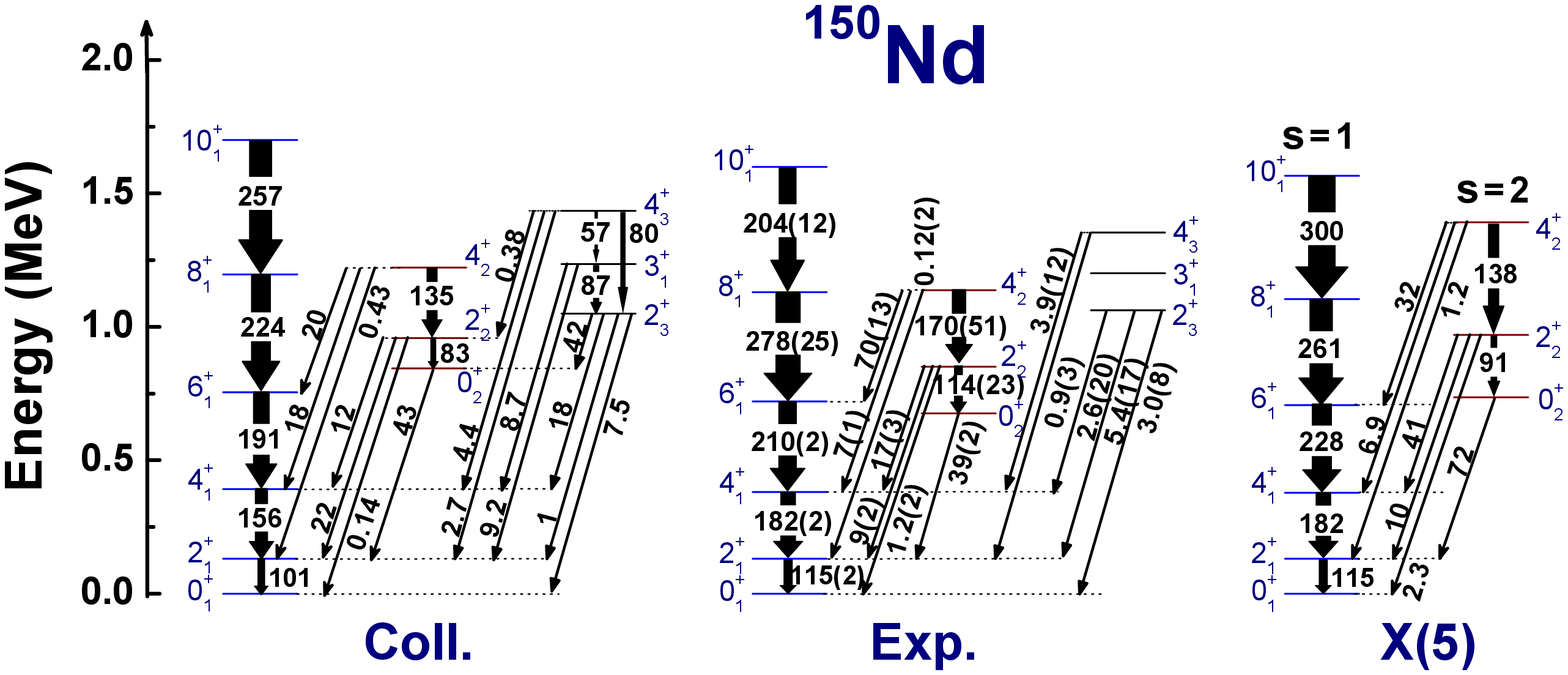}
\caption{(Color online) 
The spectrum of $^{150}$Nd calculated with the
PC-F1 relativistic density functional
(left), compared with the data \cite{Rei.02} (middle), and the 
X(5)-symmetry predictions (right) for the excitation energies, intraband 
and interband B(E2) values (in Weisskopf units) of the ground-state 
($s=1$) and $\beta_1$ ($s=2$) bands. 
The theoretical spectra are normalized to the experimental energy of the
state $2^+_1$, and the X(5) transition strengths are normalized
to the experimental B(E2; $2^+_1 \to 0^+_1$).}
\label{FigF}
\end{figure}

In Fig.~\ref{FigF} we compare the spectrum 
of the collective Hamiltonian for $^{150}$Nd with available data for positive 
parity states \cite{Rei.02,NNDC,LBNL}, and with the predictions of the 
X(5) model. For the moments of inertia of the collective Hamiltonian we 
have multiplied the Inglis-Belyaev values Eq.~(\ref{Inglis-Belyaev}) with 
a common factor determined in such a way that the calculated energy of 
the $2_1^+$ state coincides with the experimental value. In deformed 
nuclei, to a good approximation, the enhancement of the 
effective moment of inertia scales the relative excitation energies within each band 
by a common factor, but otherwise leaves the bandhead of the $\beta$ band 
($0^+_2$) unaltered. Of course, this simple relation between the scaling of 
effective moments of inertia and excitation energies within each band would be 
exact only if rotational and vibrational degrees of freedom were decoupled. 
The degree of mixing between bands can be inferred from the distribution of 
$K$-components (projection of the angular momentum on the body 
fixed-symmetry axis) of collective wave functions (cf. Table \ref{TabA}).
The transition rates are calculated in the full 
configuration space using bare charges. The inclusion of an additional scale 
parameter in our calculation was necessary because of the well known fact 
that the Inglis-Belyaev (IB) formula (\ref{Inglis-Belyaev}) predicts effective 
moments of inertia that are considerably smaller than empirical values. More
realistic values are only obtained if one uses the Thouless-Valatin
(TV) formula, but this procedure is computationally much
more demanding, and it has not been implemented in the current
version of the model. Here we rather follow the prescription of
Ref.~\cite{LGD.99} where, by comparing the TV and IB  moments of
inertia as functions of the axial deformation for superdeformed bands
in the $A=190-198$ mass region, it was shown that the
Thouless-Valatin correction to the perturbative expression IB is
almost independent of deformation, and does not include significant
new structures in the moments of inertia. It was thus suggested that
the moments of inertia to be used in the collective Hamiltonian can
be simply related to the IB values through the minimal prescription:
$\mathcal{I}_k (q) = \mathcal{I}^{IB}_k (q) (1+ \alpha)$, where $q$
denotes the generic deformation parameter, and $\alpha$ is a constant
that can be determined in a comparison with data. The value 
$\alpha \approx 0.4$ used for the excitation spectrum of $^{150}$Nd
is comparable to those determined in the mass $A=190-198$
region~\cite{LGD.99}.

When the IB effective moment of inertia is renormalized to the
empirical value, the excitation spectrum of the collective Hamiltonian determined by 
the PC-F1 density functional is in very good agreement with the available data 
for the ground-state band, (quasi) $\beta$ and $\gamma$ bands. This is also 
true for the corresponding intraband and interband B(E2) values except, perhaps, 
for the strong transition $2^+_3 \to 0^+_2$ ($\approx$ 42 Weisskopf units), not seen 
in the experiment. Such a strong transition is probably due to the mixing between 
the theoretical (quasi) $\beta$ and $\gamma$ bands. Table \ref{TabA}   
includes the distributions of $K$-components 
(projection of the angular momentum on the body fixed-symmetry axis), 
for the collective wave functions of 
the lowest three bands in $^{150}$Nd, $^{152}$Sm, and $^{154}$Gd. 
For $^{150}$Nd, in particular, we notice a pronounced mixing of the 
states $2^+_2$ and $2^+_3$, assigned to the $\beta$ and $\gamma$ bands, 
respectively. For higher angular momenta the mixing between the two bands 
is less pronounced. In addition to the results for intraband transitions, we 
notice the very good agreement between the theoretical and experimental 
value B(E2; $0^+_2 \to 2^+_1$). Again, we emphasize that the calculation 
of transition probabilities is parameter-free. Fig.~\ref{FigF} also includes the 
X(5) model predictions for ground-state ($s=1$) and $\beta_1$ ($s=2$) bands 
of $^{150}$Nd. The theoretical spectrum is normalized to the experimental 
energy of the state $2^+_1$ and, in addition, the X(5) transition strengths are 
normalized to the experimental B(E2; $2^+_1 \to 0^+_1$). The simple X(5) 
model does not reproduce the data with the same accuracy as the solution of the 
collective Hamiltonian, especially the transition rates between bands. 
This could be due to the fact that $^{150}$Nd is already slightly to 
the rotor side of the phase transition \cite{Clark.03,CZK.03}, or simply to the 
fact that the X(5) model does not include $\beta - \gamma$ 
coupling effects. Of course, 
even though the ansatz for the X(5) model involves a separation of 
variables in the $\beta$ and $\gamma$ degrees of freedom, a full set 
of predictions for the quasi-$\gamma$ band can be obtained \cite{BCZ.03}. 
However, this necessitates two additional parameters in the X(5) model 
that are adjusted, for instance, to the excitation energy of the 
quasi-$\gamma$ band and the $\Delta K = 2$ transitions. 

\begin{figure}
\includegraphics[scale=0.45,angle=0]{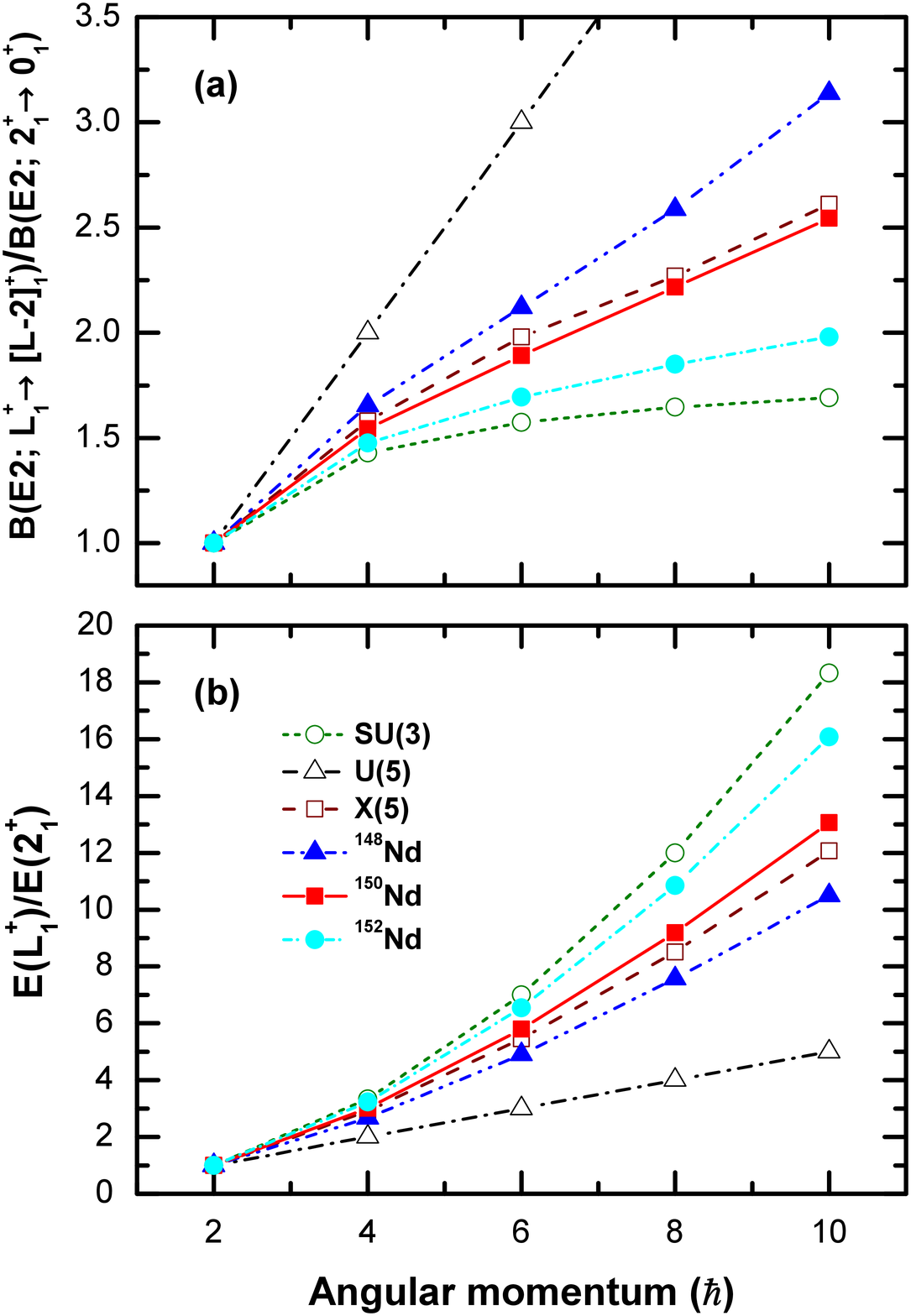}
\vspace{1cm}
\caption{(Color online)
B(E2; $L \to L-2$) values (upper panel) and 
excitation energies (lower panel) for the yrast states in $^{148}$Nd, 
$^{150}$Nd, and $^{152}$Nd, calculated with PC-F1 and compared 
with those predicted by the U(5), X(5), and SU(3) symmetries (open symbols).}
\label{FigG}
\end{figure}

When compared with our particle-number projected axial GCM calculation 
of Ref.~\cite{PRL99}, where the same PC-F1 density functional was used, 
but configuration mixing included only prolate axially symmetric wave 
functions (cf. Fig. 2 of Ref.~\cite{PRL99}), the full solution of the collective 
Hamiltonian in five dimensions displays reduced B(E2) values both for 
interband and intraband transitions, and the $\beta$ band is calculated 
at higher excitation energy. However, the inclusion of  $\beta - \gamma$ 
coupling effects does not spoil the good agreement with data. 
The comparison with the X(5) model predictions is 
also illustrated in Fig.~\ref{FigG} where, 
for the yrast states of $^{148}$Nd, $^{150}$Nd and $^{152}$Nd, 
we compare the B(E2; $L \to L-2$) values 
and excitation energies calculated using the collective Hamiltonian based 
on the PC-F1 density functional, with the corresponding values predicted 
by the U(5) and SU(3) dynamical symmetries, and the X(5) model.  
Obviously the E2 rates and excitation energies for $^{150}$Nd are closest to  
those calculated from analytic expressions corresponding to the X(5)
model. $^{148}$Nd does not differ very much from the X(5) limit, 
whereas the yrast states of $^{152}$Nd indicate that this 
nucleus is closer to a deformed rotor.   We note, for instance, 
that in the case of the axial GCM model based on the Gogny interaction,
the X(5) model predictions for the yrast excitation energies 
and transition rates were actually found to be closer to the calculated 
spectrum of $^{148}$Nd, rather than $^{150}$Nd \cite{RE.08}.
This comparison, of course, emphasizes the problem that the 
physical control parameter, i.e. the nucleon number, is not continuous and 
therefore in general a microscopic calculation cannot exactly reproduce 
the point of QPT, in contrast to phenomenological geometric or algebraic 
models that use continuous control parameters.

\begin{figure}
\includegraphics[scale=1.0]{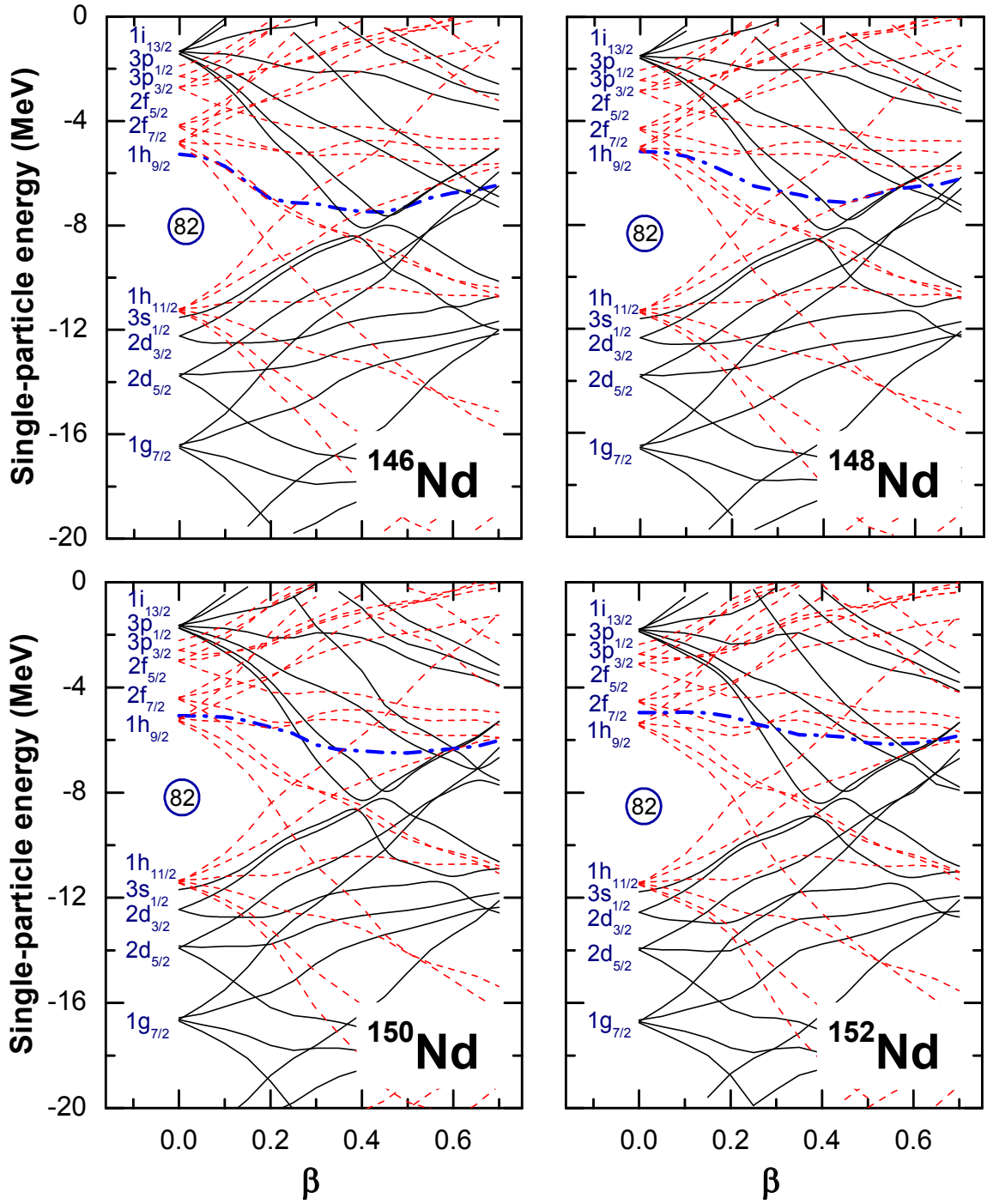}
\caption{(Color online) Neutron single-particle levels in Nd isotopes, 
as functions of the axial deformation parameter $\beta$. Thick dot-dashed 
curves denote the position of the Fermi level.}
\label{FigH}
\end{figure}

\begin{figure}
\includegraphics[scale=1.0]{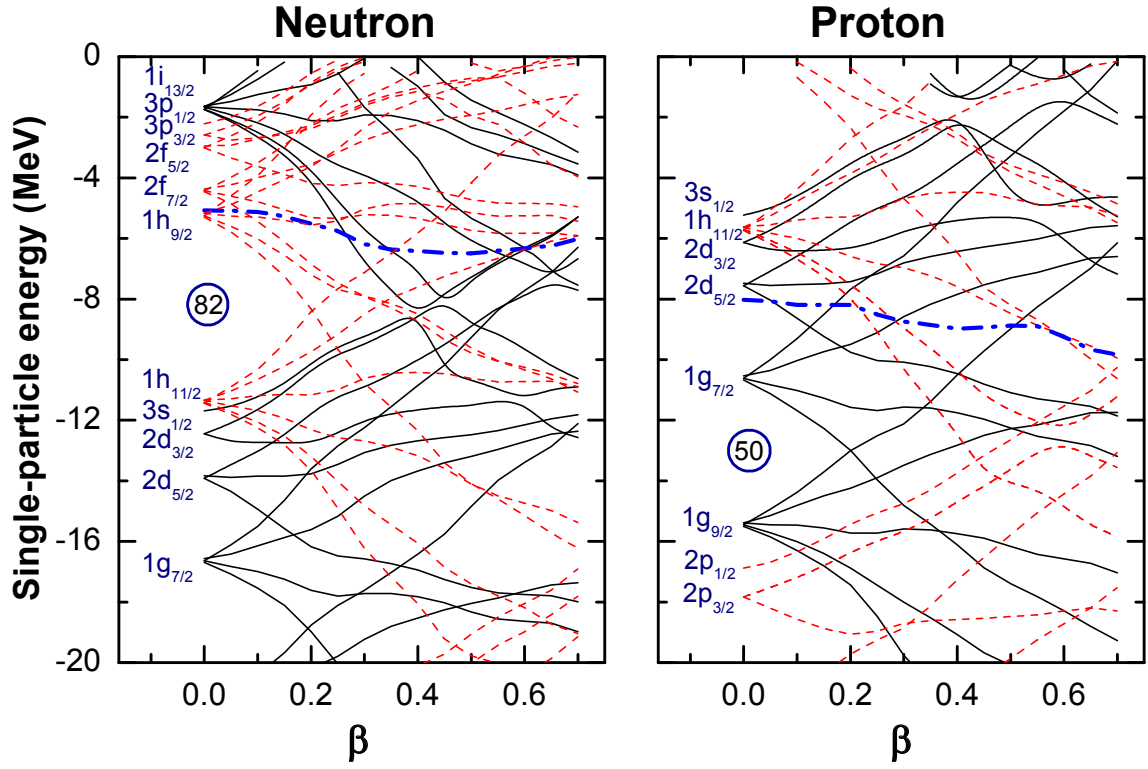}
\vspace{1cm}
\caption{(Color online) Neutron (left) and proton (right) 
single-particle levels in  $^{150}$Nd, 
as functions of the axial deformation parameter $\beta$. 
Thick dot-dashed curves denote the position of the 
corresponding Fermi levels.}
\label{FigI}
\end{figure}

A microscopic picture of the softness of the potential
with respect to $\beta$ deformation and the related phenomenon of 
QPT in $^{150}$Nd, emerges when considering the single-nucleon 
levels. In Figs.~\ref{FigH} and \ref{FigI} 
we display the neutron single-particle levels 
in $^{146-152}$Nd, and the neutron and proton single-particle levels 
in $^{150}$Nd, respectively, as functions of the axial deformation parameter 
$\beta$. The thick dot-dashed curves denote the position of the 
corresponding Fermi levels. In Fig.~\ref{FigH} we follow the evolution 
of the Fermi level with the increase in neutron number. For the range  
of deformation $0.2 \leq \beta \leq 0.4$, in particular, the Fermi level 
crosses an energy interval of low level density. In $^{150}$Nd both 
the neutron and proton Fermi levels (cf. Fig.~\ref{FigI}) are in the 
region of low level density for $0.2 \leq \beta \leq 0.4$, and this result 
in the soft-$\beta$ potential (cf. Figs. \ref{FigA} and \ref{FigB}). With 
further increase of the neutron number, the corresponding Fermi level in 
$^{152}$Nd approaches a region of high level density and the potential 
displays a pronounced, well deformed minimum at $\beta \approx 0.35$. 
Further evidence for an abrupt change of structure between 
$^{150}$Nd and $^{152}$Nd can be clearly seen in the 
experimental spectra of the neighboring odd-Z nuclei: 
$^{151}$Pm and $^{153}$Pm \cite{NNDC}. 
Qualitatively very different band structures result from the coupling of 
the odd proton to the $N=90$ or $N=92$ core nuclei. In  $^{153}$Pm 
($N=92$) one finds regular $\Delta J =1$ low-lying rotational
bands characteristic for the limit of strong coupling. On the other hand, 
the lowest bands in $^{151}$Pm ($N=90$) are regular 
$\Delta J =2$ sequences, and this indicates that the odd
proton is coupled to a core nucleus that is soft with respect to 
axial deformation.

\begin{figure}
\includegraphics[scale=0.9]{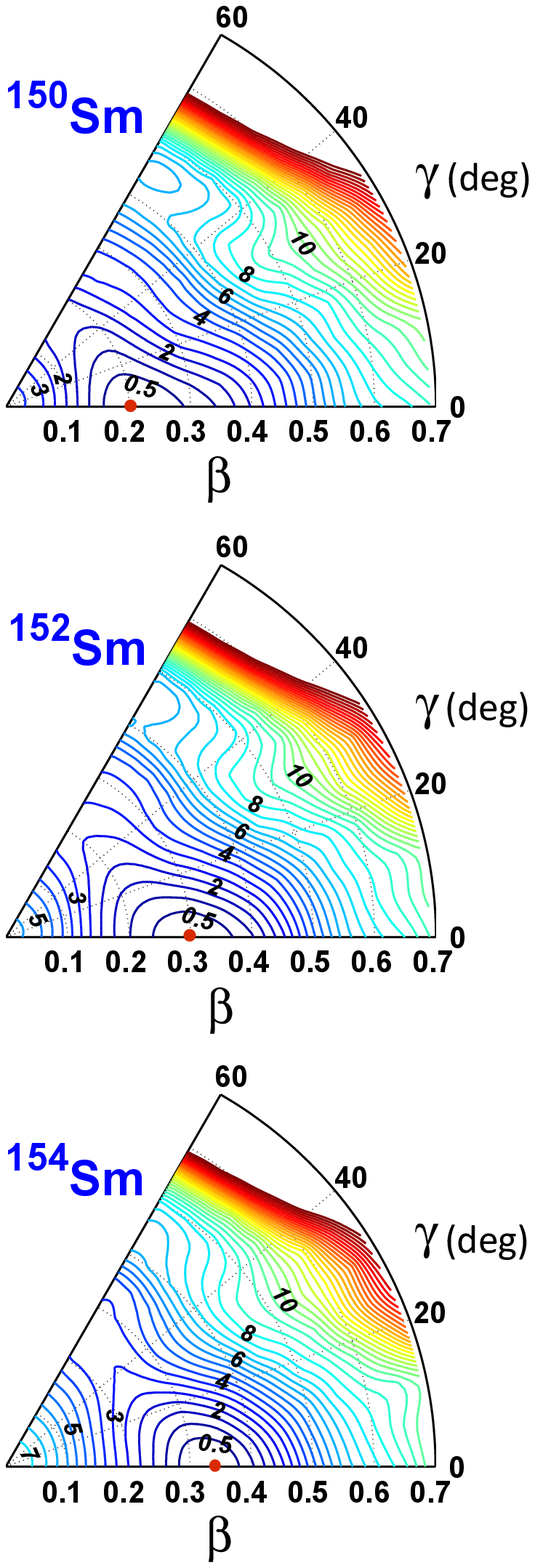}
\caption{(Color online) Self-consistent RMF+BCS triaxial quadrupole
binding energy maps of $^{150,152,154}$Sm 
in the $\beta - \gamma$ plane ($0\le \gamma\le 60^0$).
All energies are normalized with respect to
the binding energy of the absolute minimum (red filled circle). 
The contours join points
on the surface with the same energy (in MeV).}
\label{FigJ}
\end{figure}

\begin{figure}
\includegraphics[scale=0.9]{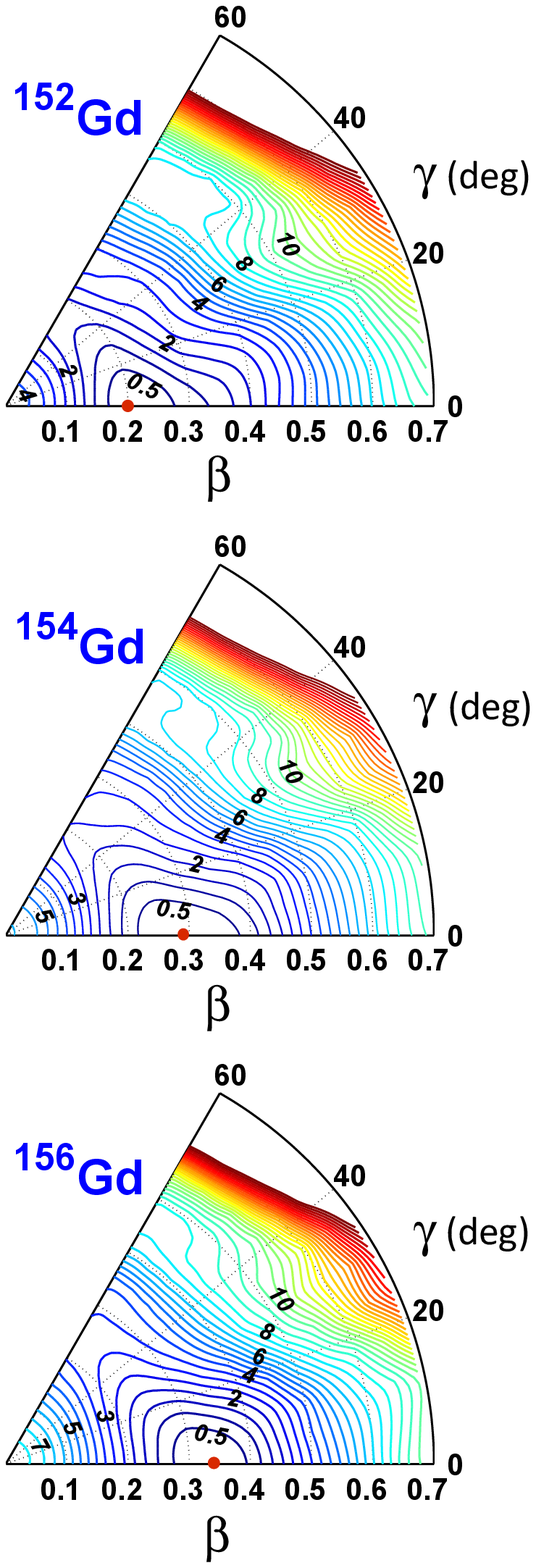}
\caption{(Color online) Same as Fig.~\ref{FigK}, but for the isotopes 
$^{152,154,156}$Gd.}
\label{FigK}
\end{figure}

The microscopic results for the Sm and Gd isotopic chains are not very different 
from those of Nd nuclei, and thus in the remainder of this section we only present 
the most interesting features related to possible QPT. The
self-consistent RMF+BCS triaxial quadrupole
binding energy maps of $^{150,152,154}$Sm 
in the $\beta - \gamma$ plane ($0\le \gamma\le 60^0$) are shown in 
Fig.~\ref{FigJ}, and those of $^{152,154,156}$Gd in Fig.~\ref{FigK}. In both 
cases we notice the increase of deformation of the prolate minima with 
increasing number of neutrons. However, in contrast to $^{150}$Nd, the 
two $N=90$ isotones do not display pronounced $\beta$-extended minima. 
This is especially the case for $^{152}$Sm, whereas a more extended 
flat prolate minimum of the potential energy surface is calculated in 
$^{154}$Gd. Note also that the model does not predict the occurrence 
of  oblate minima in the $N=92$ isotones $^{154}$Sm and $^{156}$Gd. 

\begin{figure}
\includegraphics[scale=0.75,angle=270]{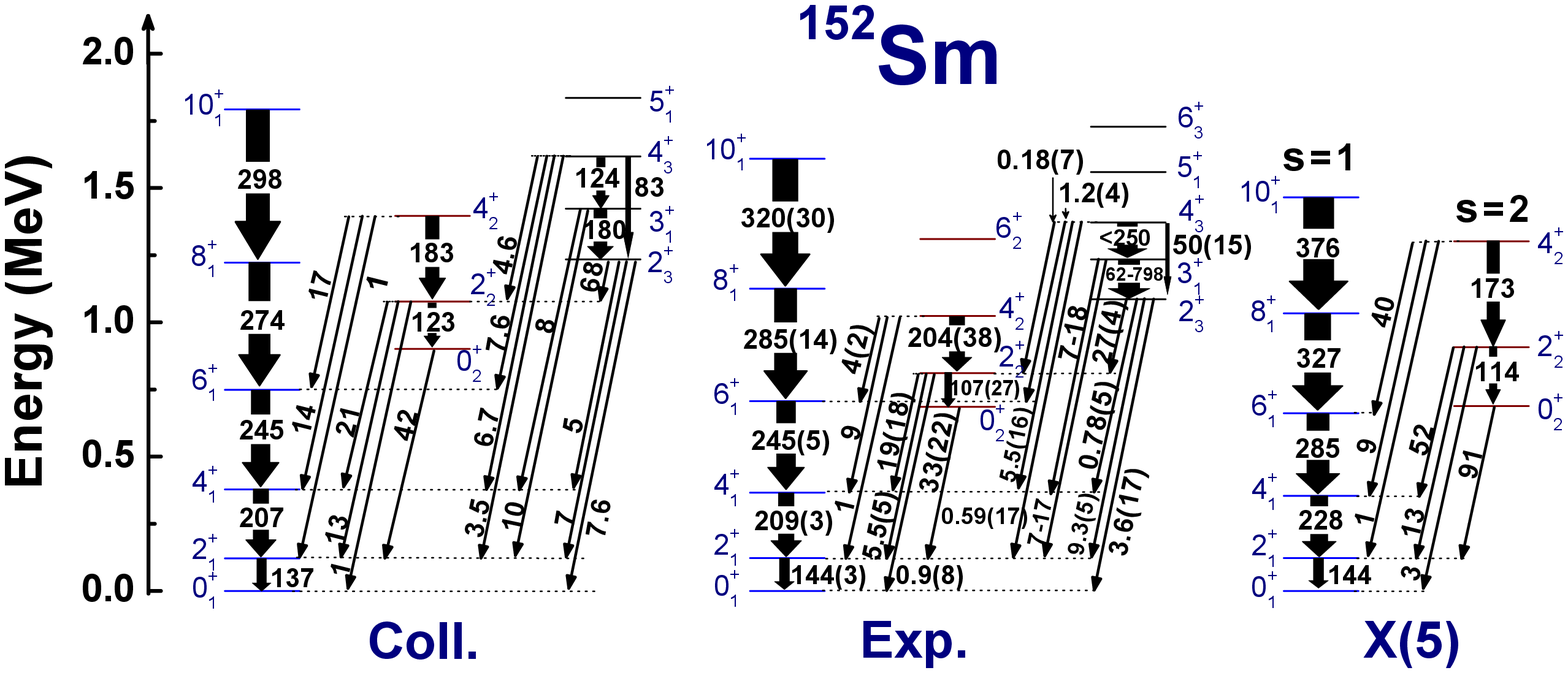}
\caption{(Color online) 
The spectrum of $^{152}$Sm calculated with the
PC-F1 relativistic density functional
(left), compared with data (middle), and the 
X(5)-symmetry predictions (right) for the excitation energies, intraband 
and interband B(E2) values (in Weisskopf units) of the ground-state 
($s=1$) and $\beta_1$ ($s=2$) bands. 
The theoretical spectra are normalized to the experimental energy of the
state $2^+_1$, and the X(5) transition strengths are normalized
to the experimental B(E2; $2^+_1 \to 0^+_1$).}
\label{FigL}
\end{figure}

\begin{figure}
\includegraphics[scale=0.75,angle=270]{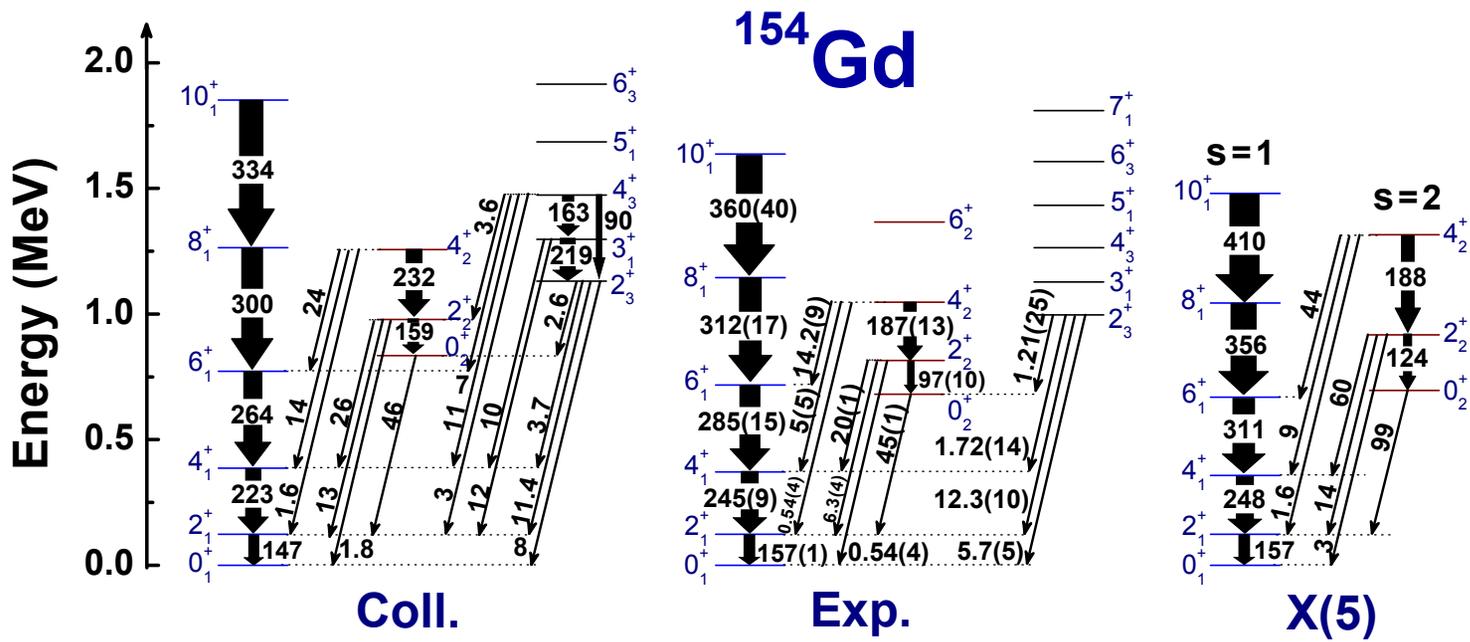}
\vspace{1cm}
\caption{(Color online) 
Same as in Fig.~\ref{FigL}, but for the
nucleus $^{154}$Gd.}
\label{FigM}
\end{figure}

\begin{figure}
\includegraphics[scale=0.45]{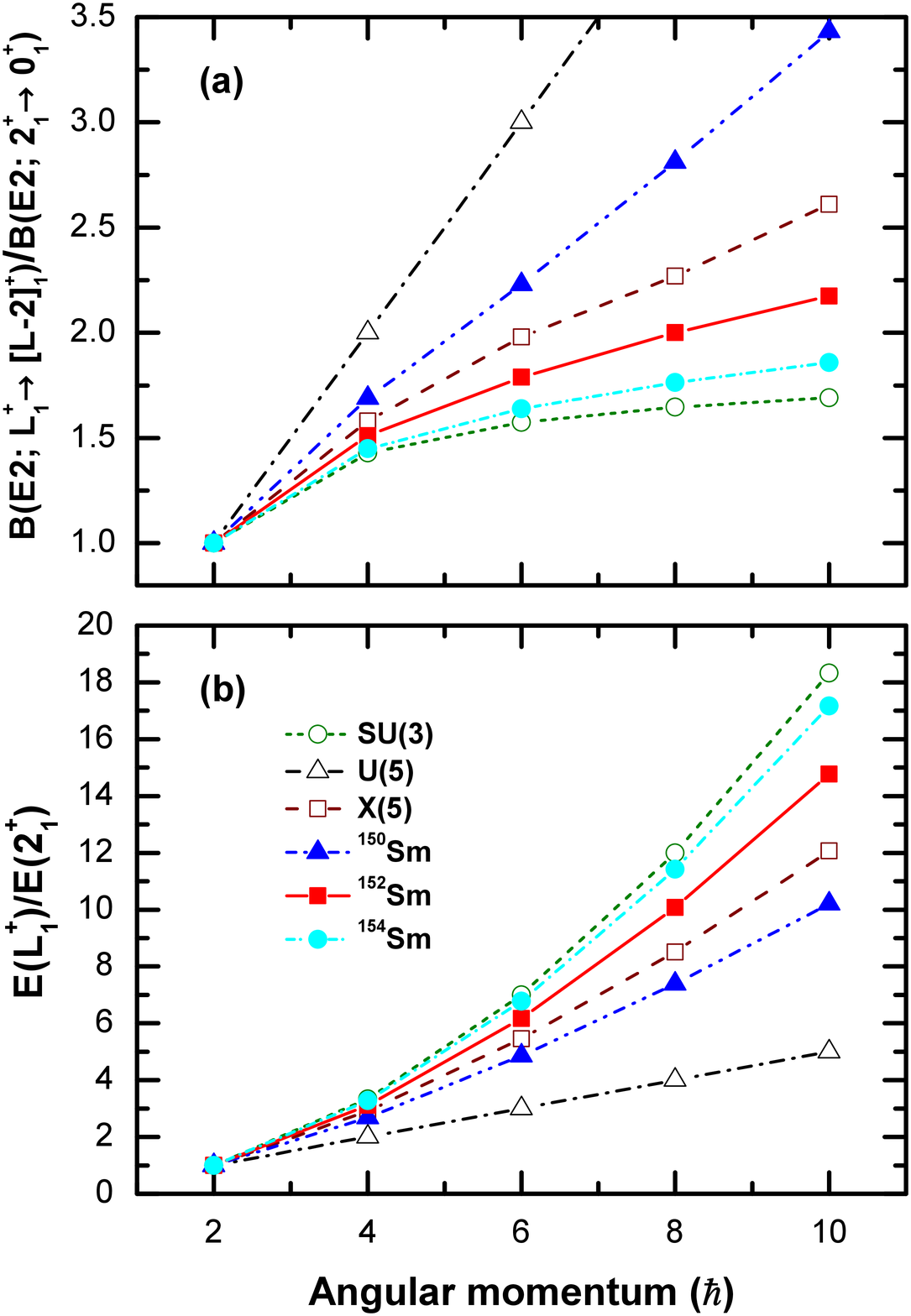}
\vspace{1cm}
\caption{(Color online) Same as Fig.~\ref{FigG}, but for the isotopes 
$^{150,152,154}$Sm.}
\label{FigN}
\end{figure}

\begin{figure}
\includegraphics[scale=0.45]{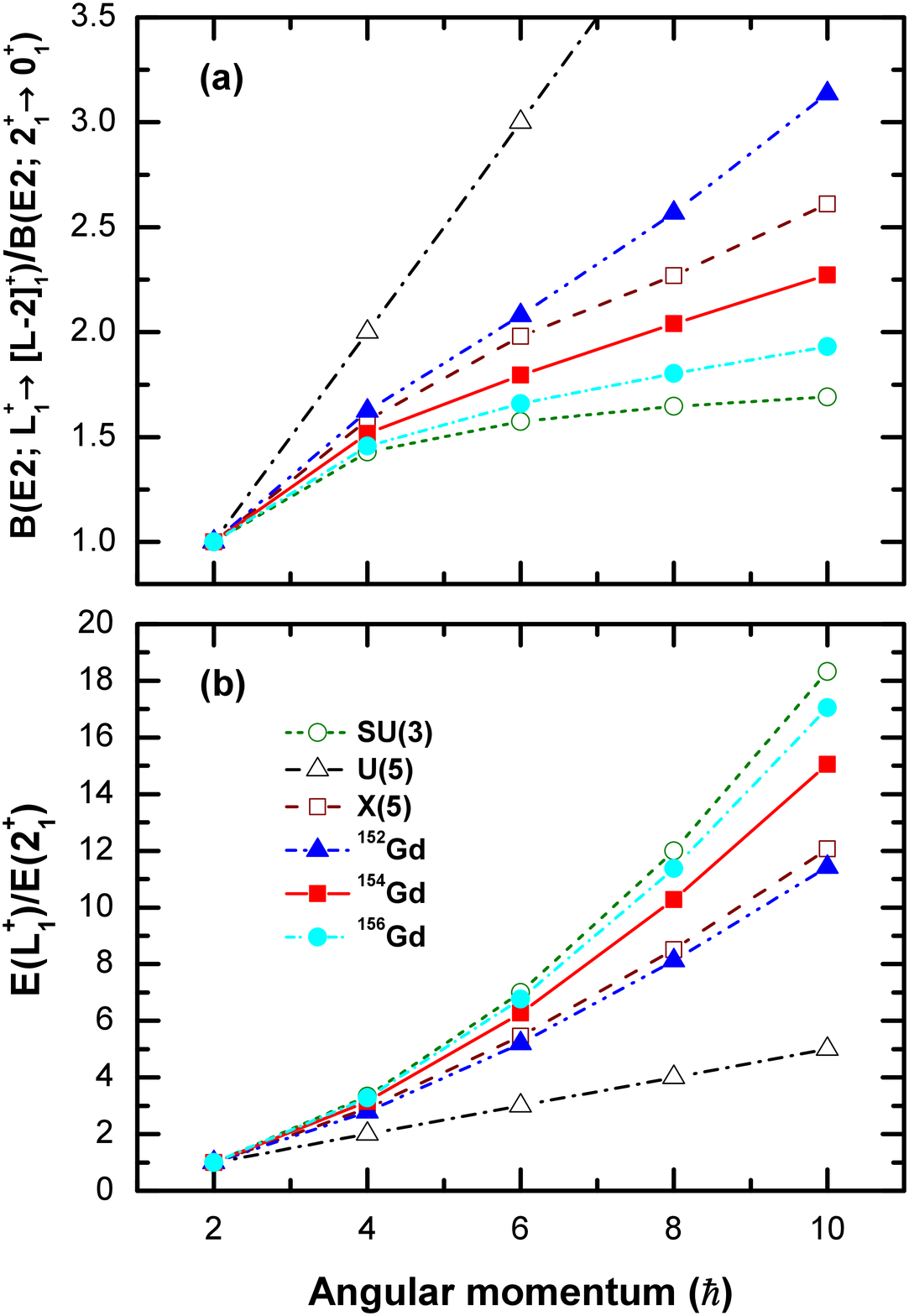}
\vspace{1cm}
\caption{(Color online) Same as Fig.~\ref{FigG}, but for the isotopes 
$^{152,154,156}$Gd.}
\label{FigO}
\end{figure}

The slight change in the potentials of the two heavier $N=90$ isotones is also 
reflected in the corresponding spectra of the collective Hamiltonian. 
Figs.~\ref{FigL} and \ref{FigM} display 
the spectra of $^{152}$Sm $^{154}$Gd, respectively, calculated with the
PC-F1 relativistic density functional, in comparison with 
data \cite{CZ.01,Ton.04,NNDC,LBNL}, 
and the X(5)-model predictions for the excitation energies, intraband 
and interband B(E2) values of the ground-state 
($s=1$) and $\beta_1$ ($s=2$) bands. For both nuclei the spectra of the 
collective Hamiltonian are in very good agreement with data, especially 
for the E2 transition rates. The agreement is perhaps not so good for the 
relative position of the $\beta$ bands, which are calculated at somewhat 
higher excitation energies compared to data, and the effective moments 
of inertia of the quasi-$\gamma$ bands are considerably smaller than the 
empirical values. As shown in Table \ref{TabA}, the mixing between the 
$\beta$ and $\gamma$ bands is much less pronounced than in the case 
of $^{150}$Nd. When compared to the predictions of the simple X(5) model, 
it is clear that the full collective Hamiltonian in five dimensions provides a more 
complete and accurate description of the low-energy spectra of
$^{152}$Sm and $^{154}$Gd, especially for interband transitions between 
the $\beta$ and ground-state bands.  The comparison with X(5) is 
further illustrated in Figs.~\ref{FigN} and \ref{FigO}, where we plot the microscopic 
B(E2; $L \to L-2$) values and excitation energies of yrast states 
of Sm and Gd $N \approx 90$ nuclei. While in the case of Nd 
isotopes the X(5) limit was very close to the results for 
$^{150}$Nd, for Sm and Gd the X(5) predictions lie in between the 
results obtained with the collective Hamiltonian for the $N=88$ and $N=90$ isotopes.  

\section{\label{secIV}Summary and Outlook}
The structural evolution of nuclei with neutron and/or proton number 
presents interesting shape transition phenomena, and 
in a number of systems signatures of first- and second-order 
quantum shape phase transitions have been observed. 
In this work we have applied the recently developed model for the 
solution of a five-dimensional collective Hamiltonian for quadrupole 
vibrational and rotational degrees of freedom,
with parameters determined by constrained self-consistent relativistic 
mean-field calculations for triaxial shapes, to a microscopic study of 
shape transitions in Nd, Sm and Gd isotopes with neutron number 
$N\approx 90$.  Available data on excitation spectra and 
transition rates indicate that the $N=90$ isotones: 
$^{150}$Nd, $^{152}$Sm, and $^{154}$Gd, are located close to the 
point of first-order QPT between spherical to axially deformed shapes.

Starting from self-consistent triaxial quadrupole binding energy maps 
in the $\beta - \gamma$ plane, calculated with the PC-F1 relativistic 
density functional, the parameters that
determine the collective Hamiltonian: mass parameters,
moments of inertia, and zero-point energy corrections, are 
calculated as functions of the deformations $\beta$ and $\gamma$. 
The diagonalization of the resulting Hamiltonian yields the
excitation energies and collective wave functions. An important 
feature of the model is that physical observables, such as E2 
transition rates, are calculated in the full configuration space. 
Using the bare value of the proton charge, parameter-free 
transition probabilities between eigenstates of the collective
Hamiltonian can be directly compared with experimental values.
In the current implementation of the model the  
moments of inertia are calculated with the Inglis-Belyaev formula. 
The resulting values are considerably smaller than 
than the empirical moments of inertia and therefore, in order to 
compare the relative excitation energies with data, it was necessary to 
increase the Inglis-Belyaev moments of inertia. The effective 
values were adjusted to reproduce the experimental energies of 
the $2_1^+$ state in the ground-state band of each nucleus. 

A detailed analysis of shape transitions in Nd isotopes has been carried 
out, from the spherical $^{144}$Nd to the strongly deformed 
$^{154}$Nd. With increasing neutron number the Nd isotopes remain 
prolate deformed in the lowest state, and the calculation reproduces 
the characteristic signatures of shape transitions: 
$R_{4/2} \equiv E(4_1^+)/E(2_1^+)$ and B(E2; $2^+_1 \to 0^+_1$), 
as well as the isotopic dependence of the effective order parameter 
$E(6^+_1)/E(0^+_2)$, which takes the special value 
of $\approx 1$ close to a first-order phase transition point.

Both the data and the results of microscopic calculations show that 
there is an abrupt change of structure at $N=90$. 
The microscopic potential energy surface of $^{150}$Nd displays 
a flat prolate minimum that extends in the interval $0.2 \leq \beta \leq 0.4$ 
of the axial deformation parameter, and a parabolic dependence on 
$\gamma$ for $\gamma \leq 30^\circ$ in the region of the flat prolate 
minimum. The resulting excitation spectrum of $^{150}$Nd, calculated with the 
PC-F1 density functional, and the corresponding intraband and 
interband B(E2) values, are in good agreement with available data 
for the ground-state band, $\beta$ and $\gamma$ bands. The calculation 
reproduces the excitation energies of the bandheads $0^+_2$  and 
$2^+_3$ of the $\beta$ and $\gamma$ bands, respectively and, when the 
Inglis-Belyaev effective moments of inertia are renormalized to the 
empirical value, also the relative excitation energies in all three bands. 
The parameter-free predictions for the B(E2)'s reproduce not only the 
experimental values for intraband transitions within the ground-state band 
and the $\beta$ band, but also for transitions from the $\beta$ and 
$\gamma$ bands to the ground-state band, including the measured 
branching ratios. The sequence of neutron single-particle 
levels in Nd isotopes, as functions 
of the axial deformation parameter $\beta$, and the isotopic 
dependence of the corresponding Fermi level, offer a qualitative 
explanation of the $\beta$-softness of the potential in $^{150}$Nd. 
Similar results have been obtained for the Sm and Gd nuclei, and in 
particular the spectra of the collective Hamiltonian and the corresponding 
E2 transition rates reproduce the data in the $N=90$ isotones 
$^{152}$Sm and $^{154}$Gd.

A number of experimental studies over the last decade have disclosed
candidate nuclei for quantum shape phase transitions in 
several mass regions. In addition to geometric and algebraic 
theoretical methods, it is also important to study these phenomena using 
microscopic models that explicitly take into account nucleonic 
degrees of freedom. Among the microscopic approaches to the 
nuclear many-body problem, the framework of nuclear energy 
density functionals (EDF) provides the most complete 
description of ground-state properties and collective 
excitations over the whole nuclide chart. This work, together with 
similar recent studies, has shown that self-consistent mean-field 
models based on the EDF framework describe not only general 
features of shape transitions, but also particular properties of spectra 
and transition rates at the critical point of QPT. However, to calculate 
excitation spectra and transition probabilities, the self-consistent 
mean-field approach must be extended to include correlations 
related to restoration of broken symmetries and fluctuations of 
collective variables. This can be done either by performing GCM 
configuration mixing calculations of projected wave functions, or 
by constructing collective Bohr-type Hamiltonians with 
deformation-dependent parameters determined from self-consistent 
mean-field calculations. The possibility to perform self-consistent 
microscopic studies of shape transitions opens a new perspective 
on the origin of nuclear QPT in various mass regions.  It is therefore 
important to systematically analyze, also employing different energy density 
functionals, various types of shape phase transitions that have 
been predicted in several regions of medium-heavy and heavy nuclei. 

\bigskip \bigskip
\leftline{\bf ACKNOWLEDGMENTS}
We thank D. Bonatsos and R. F. Casten for useful discussions.
This work was supported in part by MZOS - project 1191005-1010,
by the Major State 973 Program 2007CB815000,
the NSFC under Grant Nos. 10435010, 10775004 and 10221003,
and by the DFG cluster of excellence \textquotedblleft Origin and
Structure of the Universe\textquotedblright\ (www.universe-cluster.de).
Z. P. Li acknowledges support from the Croatian National Foundation 
for Science. The work of J.M, T.N., and D.V. was supported in part by 
the Chinese-Croatian project "Nuclear structure far from stability". 
\bigskip


\newpage

\begin{table}
\tabcolsep=6pt
\caption{\label{TabA} Distribution of the $K$-components $K=0,2,4$ 
(projection of the angular momentum on the body fixed-symmetry axis) 
in percentage (\%), for the collective wave functions of 
the lowest three bands in $^{150}$Nd, $^{152}$Sm, and $^{154}$Gd.}
\begin{center}
\begin{tabular}{c|ccc|ccc|ccc}
\hline\hline
 & \multicolumn{3}{c|}{$^{150}$Nd} & \multicolumn{3}{c|}{$^{152}$Sm} & \multicolumn{3}{c}{$^{154}$Gd}\\
\hline
        & $K=0$ & $K=2$ & $K=4$ & $K=0$ & $K=2$ & $K=4$ & $K=0$ & $K=2$ & $K=4$ \\
$2^+_1$ & 99.8& 0.2 &  0  & 99.9& 0.1 &  0  & 99.9&  0.1&  0  \\
$4^+_1$ & 99.2& 0.8 &  0  & 99.7& 0.3 &  0  & 99.6&  0.4&  0  \\
$6^+_1$ & 98.5& 1.5 &  0  & 99.3& 0.7 &  0  & 98.9&  1.1&  0  \\
$8^+_1$ & 98.0& 1.9 & 0.1 & 98.9& 1.1 &  0  & 98.1&  1.9&  0  \\
$10^+_1$& 98.0& 1.9 & 0.1 & 98.4& 1.6 &  0  & 97.0&  3.0&  0  \\
\hline
$2^+_2$ & 53.5&46.5 &  0  & 91.1& 8.9 &  0  & 96.1&  3.9&  0  \\
$4^+_2$ & 77.4&21.0 & 1.6 & 85.5&13.9 & 0.6 & 92.7&  7.1& 0.2 \\
$6^+_2$ & 79.7&18.2 & 1.7 & 80.1&19.0 & 0.9 & 89.9&  9.7& 0.4 \\
\hline
$2^+_3$ & 47.7&52.3 &  0  & 9.5 &90.5 &  0  &  4.2& 95.8&  0  \\
$3^+_1$ & 0   &100  &  0  & 0   & 100 &  0  &  0  &  100&  0  \\
$4^+_3$ & 26.4&68.2 & 5.4 & 15.5&83.0 & 1.5 &  8.1& 91.0& 0.9 \\
$5^+_1$ & 0   &95.5 & 4.5 & 0   &98.8 & 1.2 &  0  & 99.0& 1.0 \\
$6^+_3$ & 27.2&61.8 & 8.8 & 21.0&77.1 & 1.7 & 11.2& 87.0& 1.7 \\
\hline\hline
\end{tabular}
\end{center}
\end{table}

\clearpage
\clearpage
\clearpage
\clearpage
\clearpage
\clearpage
\clearpage
\clearpage
\clearpage
\clearpage
\clearpage
\clearpage
\clearpage
\clearpage
\clearpage

\end{document}